\documentclass[fleqn,usenatbib]{mnras}

\usepackage{newtxtext,newtxmath}

\usepackage[T1]{fontenc}

\DeclareRobustCommand{\VAN}[3]{#2}
\let\VANthebibliography\thebibliography
\def\thebibliography{\DeclareRobustCommand{\VAN}[3]{##3}\VANthebibliography}

\usepackage{xcolor}
\usepackage[capitalise]{cleveref}
\usepackage{url}
\usepackage{subcaption}

\usepackage{graphicx}	
\usepackage{amsmath}	

\renewcommand{\d}{\mathrm{d}}                                          
\newcommand{\diff}[2]{\frac{\mathrm{d}#1}{\mathrm{d}#2}} 

\crefname{equation}{equation}{equations}
\Crefname{equation}{Equation}{Equations}

\title[Measuring the 21-cm Dipole]{Measuring the cosmological 21-cm dipole with 21-cm global experiments}

\author[Y. D. Ignatov et al.]{
Yordan D. Ignatov$^{1}$\thanks{E-mail: yi17@ic.ac.uk (YDI)}, Jonathan R. Pritchard$^{1}$, and Yuqing Wu$^{1}$
\\
$^{1}$Department of Physics, Blackett Laboratory, Imperial College London, London SW7 2AZ, UK}

\date{Accepted XXX. Received YYY; in original form ZZZ}

\pubyear{2022}

\begin{document}
\label{firstpage}
\pagerange{\pageref{firstpage}--\pageref{lastpage}}
\maketitle

\begin{abstract}
A measurement of the 21-cm global signal would be a revealing probe of the Dark Ages, the era of first star formation, and the Epoch of Reionization. It has remained elusive owing to bright galactic and extra-galactic foreground contaminants, coupled with instrumental noise, ionospheric effects, and beam chromaticity. The simultaneous detection of a consistent 21-cm dipole signal alongside the 21-cm global signal would provide confidence in a claimed detection. We use simulated data to investigate the possibility of using drift-scan dipole antenna experiments to achieve a detection of both monopole and dipole. We find that at least two antennae located at different latitudes are required to localise the dipole. In the absence of foregrounds, a total integration time of $\sim 10^4$ hours is required to detect the dipole. With contamination by simple foregrounds, we find that the integration time required increases to $\sim 10^5$ hours. We show that the extraction of the 21-cm dipole from more realistic foregrounds requires a more sophisticated foreground modelling approach. Finally, we motivate a global network of dipole antennae that could reasonably detect the dipole in $\sim 10^3$ hours of integration time.
\end{abstract}




\section{Introduction}
The cosmological 21-cm signal, if measured, would be the best direct probe of the universe in the as-of-yet unobserved era of Cosmic Dawn, as well as the Dark Ages and Reionization. The limited information we have so far about these eras comes primarily from indirect CMB optical depth constraints \citep{2020:PlanckCollaborationAghanimAkramiVI} and observations of luminous sources with HST/JWST \citep[see e.g.][]{2022:Robertson}. Alongside upcoming experiments such as SKA that aim to map the 21-cm fluctuations, much information is to be gained in the observation of the 21-cm all-sky signal \citep{2010:PritchardLoeb}. In particular, the largest absorption trough in the theorised signal is expected to be formed as the first generation of stars emit Ly-$\alpha$ radiation into the IGM, disappearing as the IGM becomes heated by the formation of the first X-ray binaries \citep{2012:PritchardLoeb}.

\cite{2018:BowmanRogersMonsalve} announced a detection of this absorption signal by the EDGES experiment, which they found to be more than twice the depth of the largest standard cosmological predictions \citep{2017:CohenFialkovBarkana}. This has inspired a multitude of theoretical explanations beyond the standard cosmological model. The magnitude of the predicted trough can be increased by introducing an extra cooling mechanism to the baryon fluid. This may be achieved in interacting dark matter models~\citep{2018:Barkana, 2018:MunozLoeb, 2018:FraserHektorHutsi}, wherein the cooler dark matter is weakly coupled to the baryon fluid with a cross-section that favours interaction in the cosmic dark ages. Conversely, the 21-cm absorption trough may be deepened if the CMB were enhanced by a spectrally smooth excess radio background component. This component would then be fitted out in the foreground removal process. Even before the EDGES detection, the effects of such a component on the 21-cm signal were being investigated \citep{2011:FixsenKogutLevin}, following the excess radiation background detected by ARCADE 2 \citep{2011:SingalFixsenKogut}, and later LWA \citep{2018:DowellTaylor}. Both galactic (\citealt*{2020:ReisFialkovBarkana}; \citealt{2023:SikderBarkanaFialkov}) and extra-galactic sources have been considered, with \cite{2018:FraserHektorHutsi} and \cite{2018:PospelovPradlerRuderman} suggesting annihilating and oscillating DM respectively and \cite*{2019:BrandenbergerCyrShi} discussing superconducting cosmic strings. \cite{2019:MirochaFurlanetto} showed that either the enhanced cooling or enhanced radiation approaches can produce a 21-cm signal of the correct depth while agreeing with the UVLF, but could not capture the signal's exact spectral shape. \cite{2022:MittalKulkarni} found that capturing the signal shape requires either a non-conventional star formation rate density or non-standard models of stellar populations.

Alongside a claimed non-detection of the EDGES signal by the SARAS-3 experiment, with 95.3\% rejection confidence of the best-fitting EDGES trough \citep{2022:SinghJishnuSubrahmanyan}, there has been substantial discussion surrounding the validity of the EDGES data analysis methodology. Both the Galactic and extra-Galactic foregrounds are much brighter than the signal at all frequencies of interest. Foreground/signal separation in~\cite{2018:BowmanRogersMonsalve} is carried out in a manner typical to global 21-cm experiments, by assuming a signal model and relying on the spectral smoothness of the foregrounds to fit them away. \cite{2018:HillsKulkarniMeerburg} show that replacing the assumed flattened Gaussian signal model of \cite{2018:BowmanRogersMonsalve} with a sinusoide provides an equally good fit. \cite{2019:SinghSubrahmanyan} go further to demonstrate the compatibility of the EDGES data with a sinusoidally varying artefact along with a Gaussian signal. They also show that three different 21-cm templates of~\cite{2017:CohenFialkovBarkana} can equally take the place of this Gaussian. This set of analyses are troubling; very differently-shaped 21-cm signal models are able to fit the EDGES data equally well once the foreground model parameters are adjusted to accommodate this change, as the foreground models typically used are degenerate with the 21-cm model.

In order to address the uncertainty associated with the foreground removal strategy and monopole detection, we consider the simultaneous detection of the 21-cm signal's dipole component. The dipole's orientation is fixed by Earth's motion relative to the CMB rest frame, much like the CMB dipole \citep{2020:PlanckCollaborationAghanimAkramiIII}. The dipole's spectral shape is determined by the 21-cm global signal (the monopole). These stringent constraints on the nature of the dipole are unlikely to be mimicked by false troughs brought about by experimental errors, imperfectly modelled foregrounds or beam achromaticity. \cite{2017:Slosar} and \cite{2018:Deshpande} have previously identified a concurrent detection of these components as a conclusive test case of a true global 21-cm signal detection. \cite{2023:HotinliAhn} have additionally shown that measuring higher-order 21-cm multipoles may improve 21-cm global signal parameter constraints.

The state-of-the-art and upcoming global 21-cm experiments today consist of drift-scan radiometers such as EDGES \citep{2018:BowmanRogersMonsalve}, SARAS \citep{2021:NambissanT.SubrahmanyanSomashekar}, PRIZM \citep{2019:PhilipAbdurashidovaChiang}, LEDA \citep{2018:PriceGreenhillFialkov}, REACH \citep{2022:deLeraAcedodeVilliersRazavi-Ghods} and MIST \cite{2023:MonsalveByeSievers}. In this paper, we simulate such an experiment to explore the possibility of detecting the monopole and dipole simultaneously when masked by simple foregrounds consisting of a monopole and a dipole, and when masked by more realistic foregrounds based on extrapolating radio surveys. Typically, the time series data from global experiments is averaged over the entire day. We consider splitting the day into several temporal bins required to resolve the spatial properties of the dipole. We also explore the benefits of using the data from multiple experiments situated at different latitudes together to improve the dipole's detectability and break degeneracies in the inference.

This paper is organised as follows: we discuss the 21-cm dipole, introduce the foreground simulations we will consider and build up an observation model in \cref{sec:theory}. In \cref{sec:foreground_free_inference}, we examine the minimum requirements necessary to confidently detect the dipole by considering the case where the foreground contaminants have been perfectly removed from the data. We generalise the logarithmic polynomial method commonly used in 21-cm global signal extraction in \cref{sec:simple_foreground_model_inference}, and apply it to 21-cm observations contaminated by simple foregrounds. We motivate a global antenna network that would be able to extract the 21-cm monopole and dipole for these foregrounds in \cref{sec:realistic_foreground_model_inference}. We also show that the foreground removal strategy used in this work is insufficient for a 21-cm dipole detection in the presence of fully realistic foregrounds, and discuss strategies that may be successful. We conclude in \cref{sec:conclusion}, examining the necessary requirements to leverage the presence of the 21-cm dipole signal in measured data to validate a monopole detection.

\section{Theory}
\label{sec:theory}
\subsection{The 21-cm dipole signal}

\begin{figure}
    \centering
    \includegraphics[width=\columnwidth]{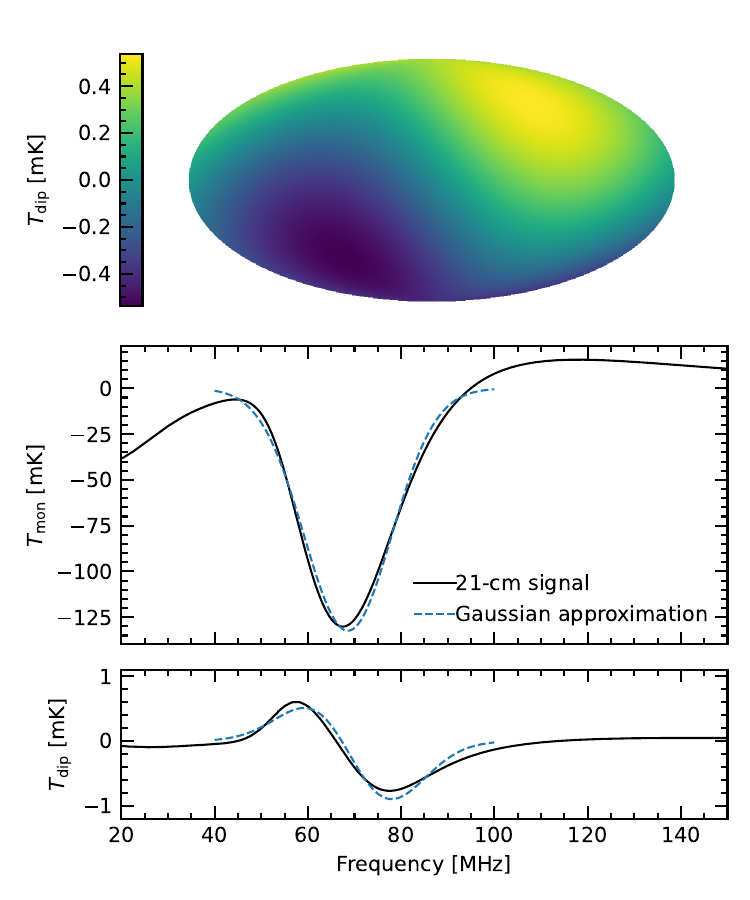}
    \caption{\textit{Top:} the 21-cm dipole sky in galactic coordinates at 60 MHz. \textit{Middle:} 21-cm sky-averaged differential brightness temperature of the 21-cm signal, generated by {\scriptsize ARES}. \textit{Bottom:} corresponding 21-cm dipole brightness temperature. The Gaussian approximation to the {\scriptsize ARES} signals is also shown (dashed lines).}
    \label{fig:ares_gaussian_fit}
\end{figure}

21-cm cosmology measures the radio signal of the 21-cm transition of neutral hydrogen. The 21-cm spin temperature, $T_S$, is defined by the relative occupation number densities of the ground $n_0$ and excited $n_1$ hyperfine states via the Boltzmann equation
\begin{equation}
    \frac{n_1}{n_0} = 3e^{-T_*/T_S} \approx 3\left(1-\frac{T_*}{ T_S}\right)\,,
    \label{eq:spintemp}
\end{equation}
where $T_*\equiv h\nu_0/k\approx 0.07$K is the characteristic temperature associated with the energy of the 21-cm transition and $\nu_0$ is the frequency of the transition. The factor of three comes from the relative degeneracy of the two states. The approximate expansion above is well-suited for all astrophysical applications, as $T_S \gg T_*$ in all cases \citep*{2006:FurlanettoOhBriggs}. The resultant 21-cm \textit{differential brightness temperature} (the difference between the 21-cm brightness temperature and CMB temperature) generated by the {\small ARES} package \citep{2014:Mirocha} running its default parameters, is shown in \cref{fig:ares_gaussian_fit} (middle panel).

The 21-cm dipole, as measured on Earth, is formed by our motion with respect to the 21-cm monopole signal present in the CMB rest frame. This motion may be separated into the Earth's orbital motion around the Sun (inducing the ``orbital dipole''), and the Sun's motion relative to the CMB in the direction $(b,l)=(48.3\degr, 264.0\degr)$, with speed $\beta\equiv v/c = 1.2\times10^{-3}$ \citep[inducing the ``Solar dipole'',][]{2020:PlanckCollaborationAghanimAkramiIII}. While Earth's orbital motion is not negligible, the 21-cm orbital dipole may be calculated and subtracted from any real-world data given the exact date and time of observation. The 21-cm dipole considered in this work therefore consists solely of the Solar dipole.

The observer's motion relative to the 21-cm signal produces two effects, the first of which is the increase in apparent brightness temperature of photons incident along the Earth's direction of motion. The second is the relativistic Doppler shift of lower frequency photons into higher frequency observation bins in the direction of motion (and vice versa in the opposite direction). The latter effect is dependent on the spectral shape of the monopole, and dominates the 21-cm dipole signal. The brightness temperature of the dipole is~\citep{2017:Slosar}
\begin{equation}
    T_\mathrm{dip}(\nu, \hat{\mathbf{n}}) = \beta \left(T_\mathrm{mon}(\nu) - \nu\diff{T_\mathrm{mon}}{\nu} \right)\hat{\mathbf{n}}\cdot \hat{\mathbf{n}}_{21} = \beta F(\nu) \hat{\mathbf{n}}\cdot \hat{\mathbf{n}}_{21} \;, \label{eq:dip_temp}
\end{equation}
where $T_\mathrm{mon}$ is the sky-averaged differential brightness temperature of the 21-cm signal, which we henceforth refer to as the monopole temperature. $F(\nu)$ is the frequency-dependent quantity in the brackets of the first equality, $\hat{\mathbf{n}}_{21}$ is the unit vector in the direction of the dipole and $\hat{\mathbf{n}}$ is the unit observation vector. The corresponding equation in \cite{2017:Slosar} contains a frequency-independent CMB dipole temperature term. The foreground model we later use treats each sky bin independently, so this frequency-independent term would be subsumed into the CMB monopole temperature. We therefore neglect it in our analysis. $T_\mathrm{dip}(\nu, \hat{\mathbf{n}})$ generated with the {\small ARES} monopole is plotted spatially in the top panel of \cref{fig:ares_gaussian_fit}, and spectrally in the bottom panel.

\subsection{Sky modelling}
\begin{figure}
    \centering
    \includegraphics[width=\columnwidth]{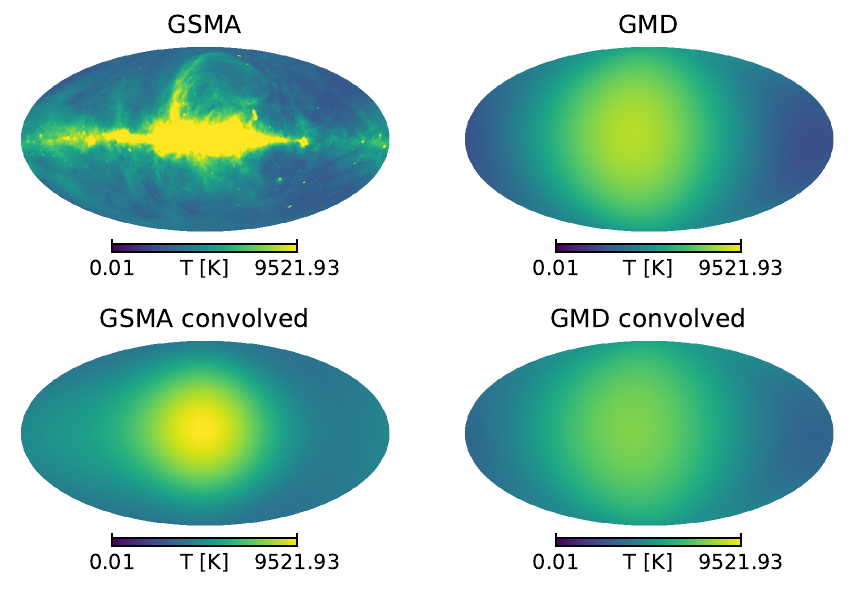}
    \caption{\textit{Top row:} Mollweide projections of the two foreground simulations considered, shown in galactic coordinates and evaluated at 60 MHz. \textit{Bottom row:} the foregrounds skies convolved with a cosine-squared beam. The pixel ranges for every sky shown are bounded by the maximum value of the convolved GSMA map for ease of comparison. The pixel temperatures of the GSMA map which exceed this value are downweighted to this maximum.}
    \label{fig:convolved_foregrounds}
\end{figure}

Throughout this paper, we will be considering several components that make up the full 21-cm sky. Namely, the 21-cm monopole, the 21-cm dipole, and the galactic and extra-galactic foreground contaminants to these signals. In order to begin simply and build complexity along the way, we model first the foreground sky as a simple monopole and dipole with power-law amplitudes in frequency, and later consider the more realistic map used in \cite{2021:AnsteydeLeraAcedoHandley}. The foreground skies are shown in the two top panels of  \cref{fig:convolved_foregrounds}, plotted using {\small HEALPIX}\footnote{{\scriptsize HEALPIX} website -- \url{http://healpix.sourceforge.net}}, and are described in further detail below.

\subsubsection{21-cm sky}
For simplicity we simulate the 21-cm monopole as the Gaussian
\begin{equation}
    T_\mathrm{mon}(\nu) = - A_\mathrm{21} \exp\left[-\frac{(\nu-\nu_\mathrm{21})^2}{2\Delta^2}\right] \,.
    \label{eq:mon_temp}
\end{equation}
This is fitted using the function \texttt{scipy.optimize.curve\_fit} to the largest predicted absorption feature of the {\small ARES} monopole temperature in \cref{fig:ares_gaussian_fit}, for the frequency range $40-100$ MHz. The resultant fiducial parameters are $A_\mathrm{21}=132.42$ mK, $\nu_\mathrm{21}=68.57$ MHz and $\Delta=9.399$ MHz. The dipole is calculated by substituting \cref{eq:mon_temp} into \cref{eq:dip_temp}. These Gaussian approximations to the monopole and dipole signals are also shown in \cref{fig:ares_gaussian_fit}.

\subsubsection{Foreground sky models}
\label{sec:Foreground Sky Models}
The Global Sky Model \citep[GSM,][]{2008:deOliveira-CostaTegmarkGaensler,2017:ZhengTegmarkDillon} is a sky simulation that interpolates between a number of radio surveys with full and partial coverage of the sky. With limited sky coverage and low-resolution data at low frequencies, the GSM suffers from discontinuous power law behaviour and negative temperatures in the frequency ranges of interest. To construct a realistic sky model more suited for this application we follow the example of \cite*{2021:AnsteydeLeraAcedoHandley} by fitting a power law index, $\gamma$, to each pixel of the 2008 GSM at the two frequencies 230 MHz and 408 MHz
\begin{equation}
    \gamma(\mathbf{\hat{n}}) = 
    \frac{
        \log\left(
            \frac{T_{230}(\mathbf{\hat{n}}) - T_\mathrm{CMB}}
            {T_{408}(\mathbf{\hat{n}}) - T_\mathrm{CMB}}
        \right)}
    {\log(230/408)} \,,
\end{equation}
where $\mathbf{\hat{n}}$ is the unit vector corresponding to the position of the pixel on the sky. The temperature of the sky in the frequency range of interest is scaled to lower frequencies following\footnote{This corrects a sign error in the corresponding equation from \cite{2021:AnsteydeLeraAcedoHandley}, which differs by a minus sign in front of the power law index.}
\begin{equation}
    T_{\mathrm{GSMA}}(\mathbf{\hat{n}}, \nu) = 
    (T_{408}(\mathbf{\hat{n}}) - T_{\mathrm{CMB}})\left(\frac
        {\nu}{408 \,\mathrm{MHz}}
    \right)^{\gamma(\mathbf{\hat{n}})}
    + T_{\mathrm{CMB}} \,.
    \label{eq:GDSMAnstey}
\end{equation}
The frequency-independent CMB present in the 2008 GSM temperature is subtracted from the temperature map before the frequency rescaling, and added back afterwards. We refer to this model as the Global Sky Model Anstey (GSMA).
\newline

For an instructive, yet simpler simulation of the 21-cm foreground sky, we use the monopole and dipole components of the GSMA sky. The foreground dipole will likely be the most problematic component to separate from the 21-cm dipole, in analogy to the foreground monopole when aiming to detect the 21-cm monopole. We fit the GSMA's monopole and dipole with amplitudes $A_\mathrm{mon/dip}$ and power law indexes $\alpha_\mathrm{mon/dip}$, as well as fitting the position of the dipole in the sky, $\hat{\mathbf{n}}_\mathrm{GMD}$. The maximum frequency-dependent dipole direction variation from the mean value was found to be $\sim 0.1\degr$, which is much smaller than the FWHM of the beam. We therefore take the dipole direction to be frequency-independent. We refer to this simulation as the Global Monopole-Dipole (GMD), given by
\begin{equation}
\begin{aligned}
    T_\mathrm{GMD}(\nu) &= A_\mathrm{mon}(\nu/60 \, \mathrm{MHz})^{-\alpha_\mathrm{mon}} \\
 &\quad+ A_\mathrm{dip}(\nu/60 \, \mathrm{MHz})^{-\alpha_\mathrm{dip}}\hat{\mathbf{n}}\cdot \hat{\mathbf{n}}_\mathrm{GMD} \,.
\end{aligned}
\end{equation}
The fitted parameters values are $A_\mathrm{mon}=5382$ K, $A_\mathrm{dip}=3100$ K, $\alpha_\mathrm{mon}=2.726$, $\alpha_\mathrm{dip}=2.555$ and the foreground dipole is centred at the galactic coordinates $(b,\,l)=(3.6\degr,\,16.3\degr)$. 

The GMD is both spectrally and spatially an approximation to the GSMA. Spatially, the GSMA and GMD are very different, with the GSMA featuring temperatures close to the galactic centre (down-weighted in \cref{fig:convolved_foregrounds}) that are hundreds to thousands of times brighter than the GMD's maximum amplitude. However, the broad beam of a dipole antenna experiment smears out the concentrated power of the GSMA onto larger scales. The convolution of both foreground simulations with a cosine-squared beam are shown in the two bottom panels of \cref{fig:convolved_foregrounds}. The convolved GSMA's maximum amplitude is still slightly brighter than that of the convolved GDM, owing to the contribution from higher-order multipoles.

Spectrally the GSMA is highly complex, featuring over 3 million pixels, each with their own spectral index. A sky observation with a wide-angle beam takes a weighted sum of many of these indices. The resultant temperature may be modelled as a power law with a number of running terms. The GMD only contains two distinct power law indices, so the number of running terms required to fit the observed temperature across multiple frequencies is smaller.

\subsection{Observation model}
\label{sec:skob_21}
To simulate observations of the sky, we neglect ionospheric effects and assume a perfectly axis-symmetric, non-chromatic beam. Our choice of beamfunction $B$ is the Hertzian dipole as a functionally simple, idealised beam \citep{1986:RybickiLightman}
\begin{equation}
    B(\hat{\mathbf{n}}; \hat{\mathbf{n}}'(t)) = \begin{cases}
        (3/2\pi) \, |\hat{\mathbf{n}} \cdot \hat{\mathbf{n}}'(t)|^2 & \text{for } \hat{\mathbf{n}} \cdot \hat{\mathbf{n}}'(t) > 1 \, \\
        0 & \text{otherwise } 
    \end{cases}\;,
    \label{eq:beamfunction}
\end{equation}
where $\hat{\mathbf{n}}'(t)$ is the unit vector direction that the beam instantaneously points in. Here, $t$ denotes \textit{local sidereal time}, and is simply a coordinate with a period of 24 hours.
The normalisation factor ensures that the beam integrates to unity over the sky. A drift-scan experiment produces continuous time-series data, which is binned into several bins per day. As the Earth rotates, within the $j^\mathrm{th}$ bin the beamfunction smears across the sky, producing the \textit{smeared beamfunction}
\begin{equation}
    B_j(\hat{\mathbf{n}}) = \int_{t_j}^{t_j+\Delta t} \frac{\d t}{\Delta t} \; B(\hat{\mathbf{n}};\hat{\mathbf{n}}'(t)) 
    \approx \frac{1}{N_p}\sum_{i=1}^{N_p}  B(\hat{\mathbf{n}};\hat{\mathbf{n}}'_{i,j})\,,
    \label{eq:time_int}
\end{equation}
where the $j^\mathrm{th}$ bin ranges from $t=t_j$ to $t=t_j + \Delta t$. We have approximated the integral as a Riemann sum over a set of $N_p$ distinct pointings per bin, where $\hat{\mathbf{n}}'_{i,j} \equiv \hat{\mathbf{n}}'(t_j + i \Delta t / N_p)$. This approximation works well owing to the broad observation beam. The sky temperature observed in the $j^\mathrm{th}$ bin by a normalised smeared beamfunction is generally given by
\begin{equation}
    T^\mathrm{obs}_j(\nu) = \int \d \hat{\mathbf{n}} \, B_j(\hat{\mathbf{n}}) \, T_\mathrm{sky}(\hat{\mathbf{n}}, \nu) \,.
    \label{eq:observation_integral}
\end{equation}
We can calculate the observed 21-cm temperature in the $j^\mathrm{th}$ bin by substituting \cref{eq:dip_temp,eq:beamfunction,eq:time_int} into \cref{eq:observation_integral}
\begin{equation}
    T^\mathrm{obs}_{21,j}(\nu) = T_\mathrm{mon}(\nu) + \frac{3}{4}\beta F(\nu) \, \frac{1}{N_p}\sum_{i=1}^{N_p} \hat{\mathbf{n}}_{i,j}'\cdot\hat{\mathbf{n}}_{21} \,,
    \label{eq:observed_mondip}
\end{equation}
where the monopole term is unchanged for a normalised smeared beamfunction, and the second term is the observed dipole temperature in each bin, $T_{\mathrm{dip},j}^\mathrm{obs}(\nu)$.
Similarly, the observed GMD temperature in the $j^\mathrm{th}$ bin is
\begin{equation}
\begin{aligned}
    T_\mathrm{GMD,j}^\mathrm{obs}(\nu) &= A_\mathrm{mon}(\nu/60 \, \mathrm{MHz})^{-\alpha_\mathrm{mon}} \\
    &\quad + \frac{3}{4} A_\mathrm{dip}\left(\frac{\nu}{60\,\mathrm{MHz}}\right)^{-\alpha_\mathrm{dip}}\frac{1}{N_p}\sum_{i=1}^{N_p} \hat{\mathbf{n}}'_{i,j}\cdot \hat{\mathbf{n}}_\mathrm{GMD}\,. 
\label{eq:observed_fgmondip}
\end{aligned}
\end{equation}
\newline

Unlike the 21-cm signal sky and the GMD, the GSMA is defined as a piecewise function on the sky for each sky pixel. Using {\small HEALPIX}, we calculate the smeared beamfunction sky maps and evaluate \cref{eq:observation_integral} for $T_\mathrm{sky}=T_\mathrm{GSMA}$, which takes the form of a sum over all sky pixels.
\newline

To generate noisy observations of the desired sky components, we first generate noise-free observations in each frequency and sky bin in a vector: $T^\mathrm{obs}_j(\nu_k)=\{T^\mathrm{obs}_1(\nu_1), T^\mathrm{obs}_1(\nu_2), ..., T^\mathrm{obs}_2(\nu_1), ...\}$. Noisy observations are then sampled from a $N_\mathrm{freq}\times N_\mathrm{bin}$-dimensional Normal distribution with mean $T^\mathrm{obs}_j(\nu_k)$ and variance $\sigma_j(\nu_k)^2$. The variance is calculated using the radiometer equation
\begin{equation}
    \sigma_j(\nu) = \frac{T^\mathrm{obs}_{j}(\nu)}{\sqrt{ t_\mathrm{int}\,\Delta \nu}} \,,
    \label{eq:radiometer_eq}
\end{equation}
where $t_\mathrm{int}$ is the integration time for the $j^\mathrm{th}$ bin and $\Delta \nu$ is the bandwidth of each frequency bin. The total integration time is then equal to $t_\mathrm{int} \times N_\mathrm{freq} \times N_\mathrm{bin}$. Since the sky temperature is dominated by foregrounds, we neglect the contribution of the observed 21-cm temperature to $T^\mathrm{obs}_{j}$, and take it to equal the observed temperature of the foreground simulation used.

\begin{figure*}
    \centering
    \includegraphics[width=0.8\textwidth]{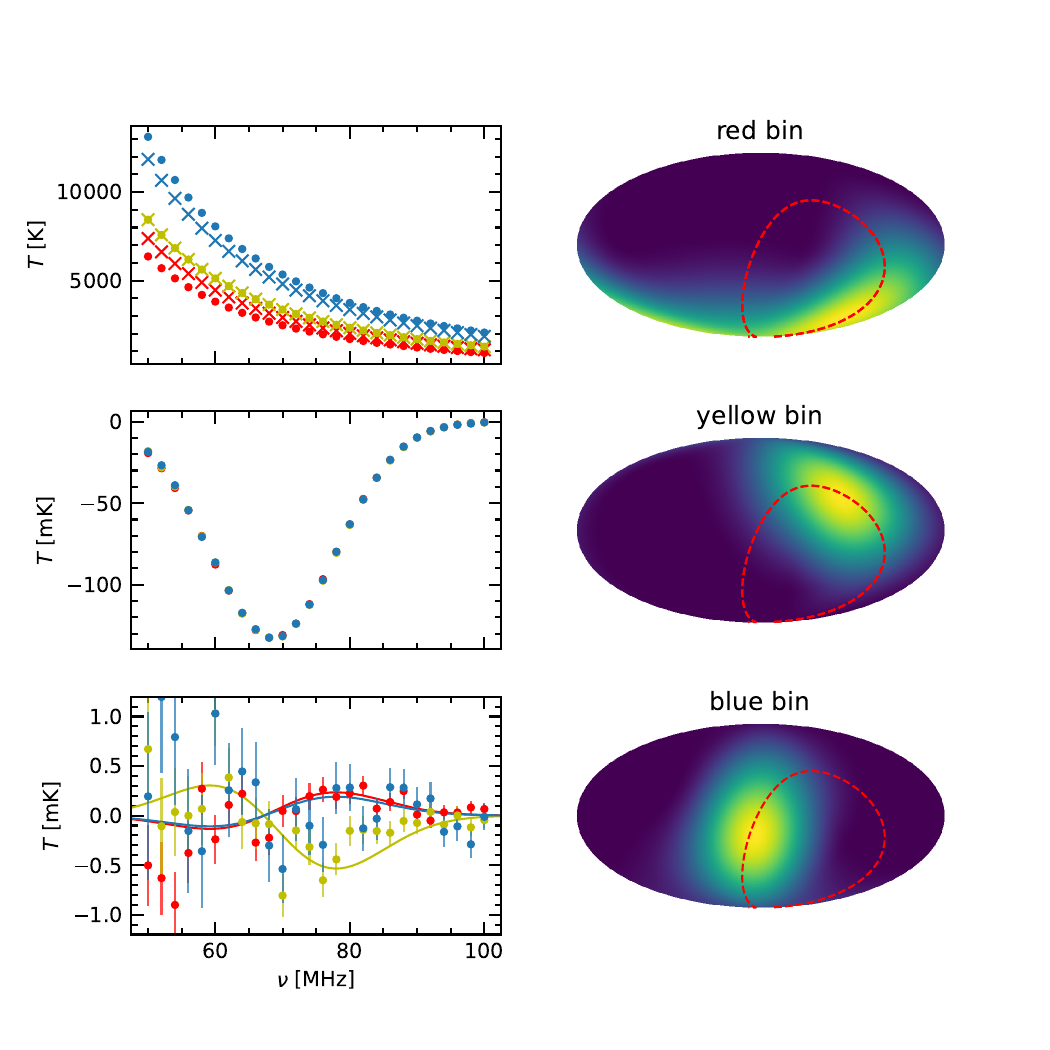}
    \caption{Single-antenna mock observations of the radio sky across three equally-spaced time bins over a 24-hour period, with an integration time of $10^5$ hours and 2 MHz frequency bins. \textit{Left column:} we plot the full observation (top) for the GSMA as dots and the GMD as crosses, the observation with foregrounds removed (middle), and the observation with foregrounds and 21-cm monopole removed (bottom). The noise-free dipole temperatures for each bin are also shown as solid lines. \textit{Right column:} Mollweide projection of the smeared beamfunctions used to observe the sky for each time bin in galactic coordinates. The red dashed line represents the direction of the zenith tracked out by the antenna (located at the MRO site) as the Earth rotates.}
    \label{fig:demonstrate_model}
\end{figure*}

Noisy observations of both foreground simulations, the 21-cm monopole and 21-cm dipole are shown in \cref{fig:demonstrate_model}, for a single-antenna dipole experiment situated at the Murchison Radio-astronomy Observatory (MRO) site in Western Australia, at a latitude of $-26.7\degr$. The three smeared beamfunctions used to observe the sky temperatures in each of three equally-spaced time bins are also shown. The temperatures are measured in 2 MHz frequency bins, for a total integration time of $10^5$ hours. A value of $10^5$ hours is chosen here to produce a noise level in each bin comparable to (but smaller than) the peak dipole amplitude in each of the time bins of \cref{fig:demonstrate_model}. This choice produces a noise level at 78 MHz (the peak of the observed 21-cm dipole) averaging 0.18 mK across the three time bins, while the fiducial dipole signal amplitude averages 0.32 mK. In \cref{sec:multi-antenna inference}, we find that this integration time is a reasonable order-of-magnitude estimate of the time required to detect the dipole in the presence of foregrounds. For these reasons, $10^5$ hours is used repeatedly throughout this work.

Although the antenna beam spreads the power of the GSMA onto larger scales, it does the same to the GMD. Therefore the observed GSMA temperatures are still greater than GMD temperatures in sky regions closer to the galactic centre (blue bin in \cref{fig:demonstrate_model}), and vise-versa away from it (red bin).

\section{Foreground-Free Inference}
\label{sec:foreground_free_inference}
We begin our analysis by considering the 21-cm signal alone, assuming that a perfect foreground removal method has been applied to the data. We generate noise assuming GSMA foregrounds. Inference is carried out using the {\small EMCEE} package \citep{2013:Foreman-MackeyHoggLang}.

\begin{figure}
    \centering
    \includegraphics[width=\columnwidth]{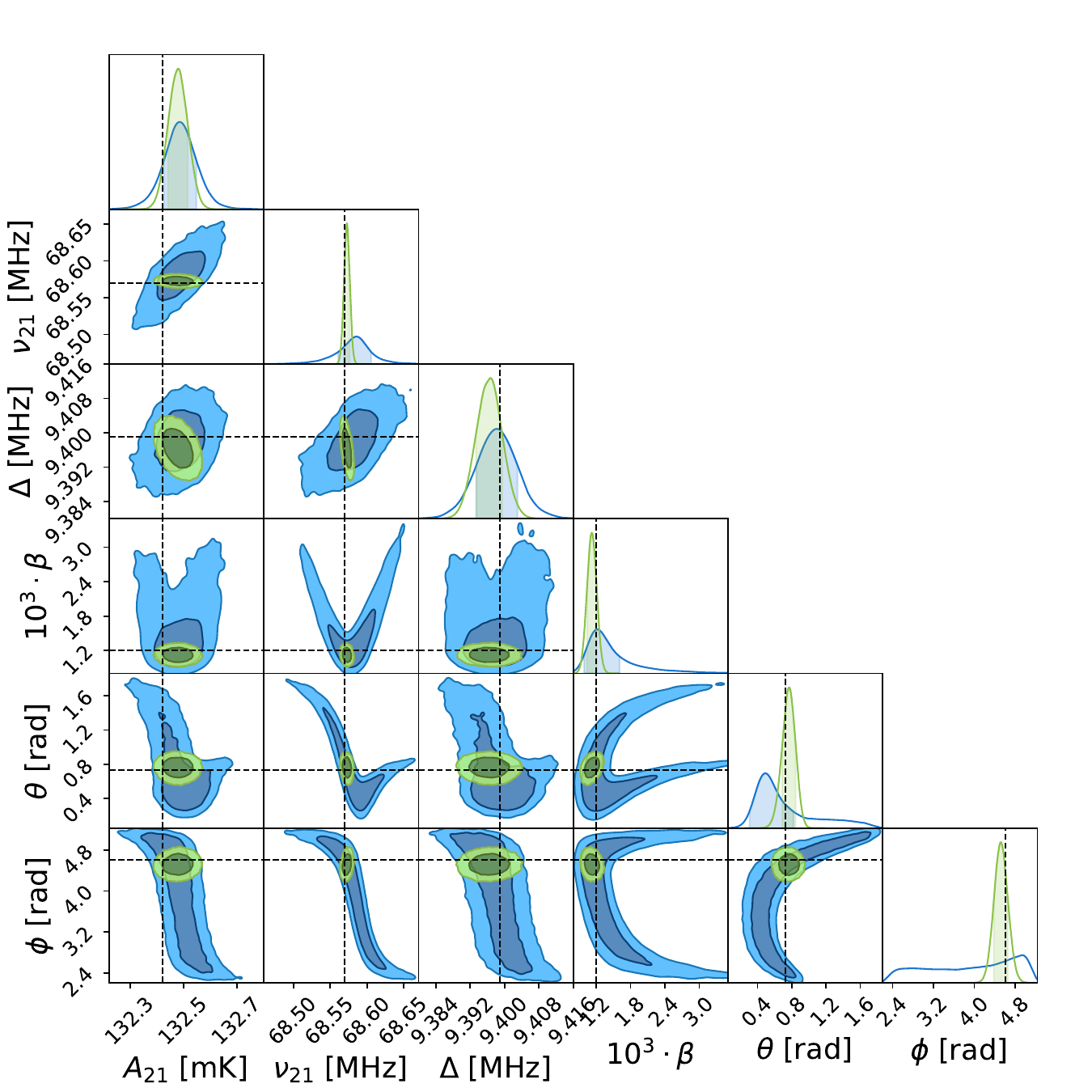}
    \caption{Posteriors of the 21-cm monopole and dipole inference of single-antenna observations (blue) and two-antenna observations (green), both for $10^5$ hours of integration time and 3 time bins per antenna. The fiducial parameter values are shown (dashed lines). The first three parameters correspond to the spectral shape of the Gaussian monopole, while the latter three describe the amplitude and direction of the dipole.}
    \label{fig:cm21_only_inference}
\end{figure}

\begin{figure}
        \centering
        \begin{subfigure}{\columnwidth}
            \includegraphics[width=\columnwidth]{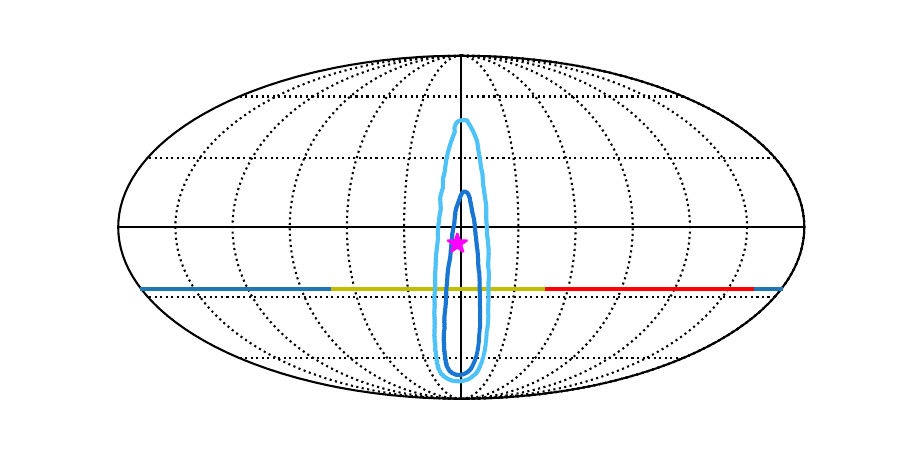}
        \end{subfigure}
        \hfill
        \begin{subfigure}{\columnwidth}
            \includegraphics[width=\columnwidth]{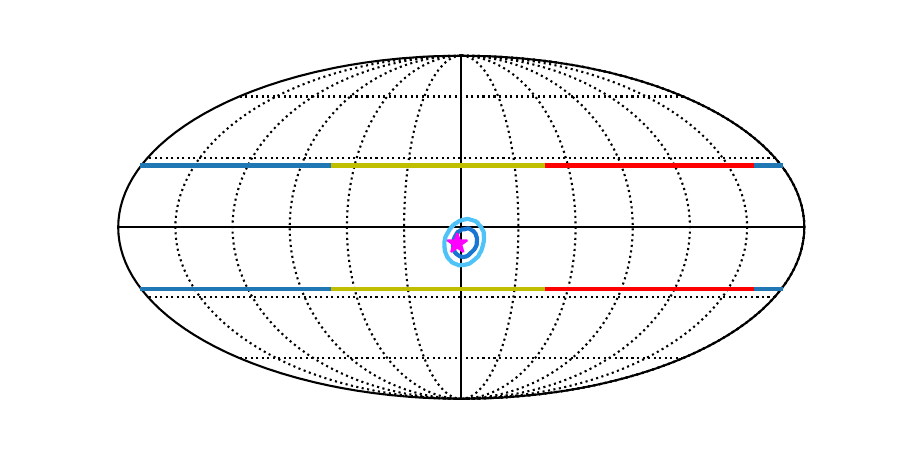}
        \end{subfigure} 
        \caption{Mollweide projection in equatorial coordinates of the marginalised posteriors of the 21-cm dipole direction, inferred from observations of the 21-cm sky using a single antenna (top) and two antennae (bottom). Mock data was generated with $10^5$ hours of integration time. The fiducial dipole direction is denoted by a star. The drift-scan antenna tracks are shown as horizontal lines, with colours denoting the longitude boundaries of the time bins.}
        \label{fig:breaking_degeneracy_projplot}
\end{figure}

\subsection{Single-antenna inference}
\label{sec:fg_free_inf}
We consider the inference of the 21-cm parameters from observations of a single dipole antenna experiment situated at the MRO site, chosen to simulate the sky seen by the EDGES experiment.

At least three time bins are a necessary requirement to infer the location of the 21-cm dipole in the sky. The Gaussian monopole parameters and the dipole magnitude parameter affect both the spectral and spatial structure of the 21-cm sky. Meanwhile, the dipole direction parameters affect only the spatial structure. If only the dipole were present in the sky, at least three time bins would be required (one for each degree of freedom of the dipole) to accurately infer the location of the dipole. The remaining three parameters, corresponding to the spectral shape of the monopole, are constrained using the spectral information of the observations. 

When the 21-cm monopole is added to the sky, the dipole parameters become degenerate with themselves and with the monopole parameters. This is seen the corner plot of \cref{fig:cm21_only_inference}, showing the posterior distribution of the 21-cm monopole and dipole inference in three time bins. Note that the number of model parameters do not change with the addition of the 21-cm monopole; consequently, the cause of the degeneracy is the presence of the monopole signal itself, intermingling with the dipole signal. A total integration time of $10^5$ hours is again used here.

The top panel of \cref{fig:breaking_degeneracy_projplot} shows the posterior dipole location of \cref{fig:cm21_only_inference} in an equatorial projection. The degenerate region lies along a line perpendicular to the drift-scan track of the antenna, which is colour-coded to represent the three time bins. The fiducial location of the dipole is intersected by this degenerate region. The use of a larger number of time or frequency bins does not further collapse the uncertainty region. Together, these findings indicate that the degeneracy is formed by the ability to move the dipole location perpendicularly away from the observation track, while changing a combination of the other parameters, in such a way as to keep the temperature of the sky viewed from any point of the observation track at a fixed temperature. The nature of the degeneracy suggests the need for multiple antennae at different Earth locations to properly localise the dipole location.

Despite the degeneracy, the monopole parameters in \cref{fig:cm21_only_inference} are well-constrained and the presence of the dipole is detected, as a non-zero $\beta$ is inferred. A non-zero dipole magnitude certainly provides useful evidence for the validation of a 21-cm monopole detection. However, a full cross-check would involve matching the 21-cm dipole direction with the CMB dipole.

\subsection{Multi-antenna inference}
\label{sec:two_antenna}
In order to collapse the degeneracy between the 21-cm dipole parameters for observations of the foreground-free 21-cm sky, we simulate observations from a second antenna. This antenna (antenna 2) is situated at a latitude of $26.7\degr$ (negative the latitude of the MRO). This is an arbitrary choice, and other reasonably-spaced latitude pairs produce similar results. Both antenna's time bins are taken to begin and end at the same longitudes. This is equivalent to having two antennae at the same longitude, with time bins that begin and end at the same LST. The temperatures measured by antenna 2 are appended to the observation vector $T^\mathrm{obs}_j(\nu_k)$, and \cref{eq:observed_mondip} is used to fit them, with smeared beamfunctions calculated for the two different antenna latitudes.

The addition of these measurements removes the 21-cm dipole parameter degeneracy of \cref{sec:fg_free_inf}, as seen in  \cref{fig:cm21_only_inference}. The bottom panel of \cref{fig:breaking_degeneracy_projplot} shows the posterior distribution of the 21-cm dipole direction projected in equatorial coordinates, which has collapsed onto a single point with the addition of the second antenna track. The two antenna's time bins are colour-coded by longitude, so appear as three distinct colours, despite there being six distinct bins in total. The extra antenna's observations additionally double the integration time, resulting in a dipole location that is better constrained laterally (parallel to the observation track).
\newline

\begin{figure}
    \centering
    \includegraphics[width=\columnwidth]{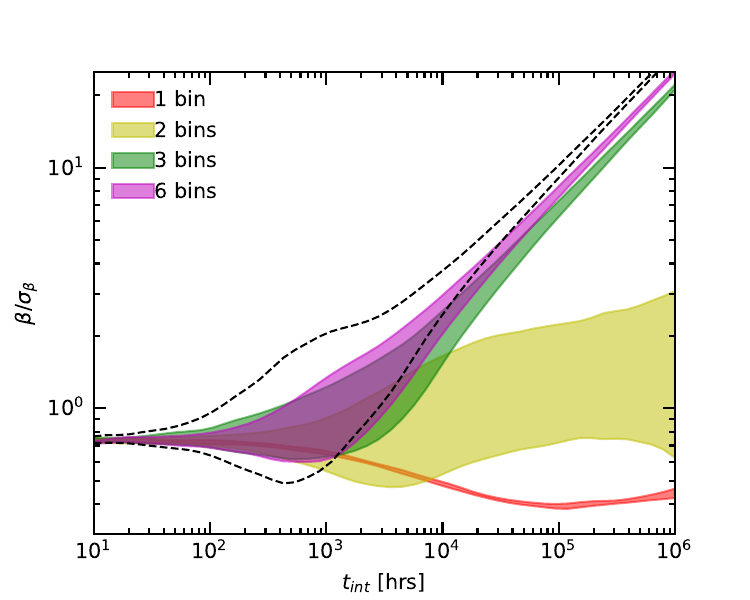}
    \caption{$1\sigma$ confidence intervals of the detection SNR of the 21-cm dipole magnitude parameter $\beta$, plotted as a function of total integration time for two antennae. This is evaluated for a number of different time bins. In the two-bin case, the different possible bin longitudes are averaged over. Additionally shown (dashed lines) is the corresponding $1\sigma$ confidence interval for four antennae and six time bins.}
    \label{fig:2ant_beta_SNR}
\end{figure}

We test the ability of different two-antenna observation strategies to detect the 21-cm dipole as a function of integration time, by making an estimate for the SNR of $\beta$, shown in~\cref{fig:2ant_beta_SNR}. The integration time is calculated as a sum over all frequency bins and time bins. To calculate this for each integration time, we find the marginalised posterior of $\beta$ inferred from 10 observations of the 21-cm sky, each with a different noise realisation. For each of the 10 posteriors, the SNR estimate is computed as the mean of the distribution, divided by the 1$\sigma$ width. For each integration time, the 10 SNR values are used to compute a $1\sigma$ confidence interval. This gives us a lower bound estimate of the integration time required to detect the monopole and dipole simultaneously with two antennae, assuming a perfect foreground removal method.

We find that an integration time between $10^4$ and $10^5$ hours is required to confidently detect the dipole. Three time bins are required as a minimum, and more than that make little difference to the SNR. Two time bins are also able to detect the dipole, but must be oriented with the dipole in close longitude alignment with the bin centre. The variability in SNR is shown in~\cref{fig:2ant_beta_SNR}, where the 2-bin case has been averaged over all possible time bin orientations. 

Also shown in~\cref{fig:2ant_beta_SNR} is the dipole detection SNR for 6 time bins and 4 antennae, situated at latitudes $-53.5\degr$, $-26.7\degr$, $26.7\degr$, and $53.5\degr$. The detection SNR improves relative to the 6-bin, 2-antenna case in proportion to the increased integration time that doubling the number of antennae offers. This suggests a solution to the large integration time required to confidently detect the dipole: a global network of many simple dipole antenna experiments. This possibility is further discussed in~\cref{sec:realistic_foreground_model_inference}.

No significant difference was found between using 1, 2 or 5 MHz frequency bins. This is likely due to there being only five parameters in the model which impact the spectral structure of the signal, which is much smaller than the number of frequency bins. For the rest of this work, we use 2 MHz bins.

\section{Simple Foreground Inference}
\label{sec:simple_foreground_model_inference}

\begin{figure}
    \centering
    \includegraphics[width=\columnwidth]{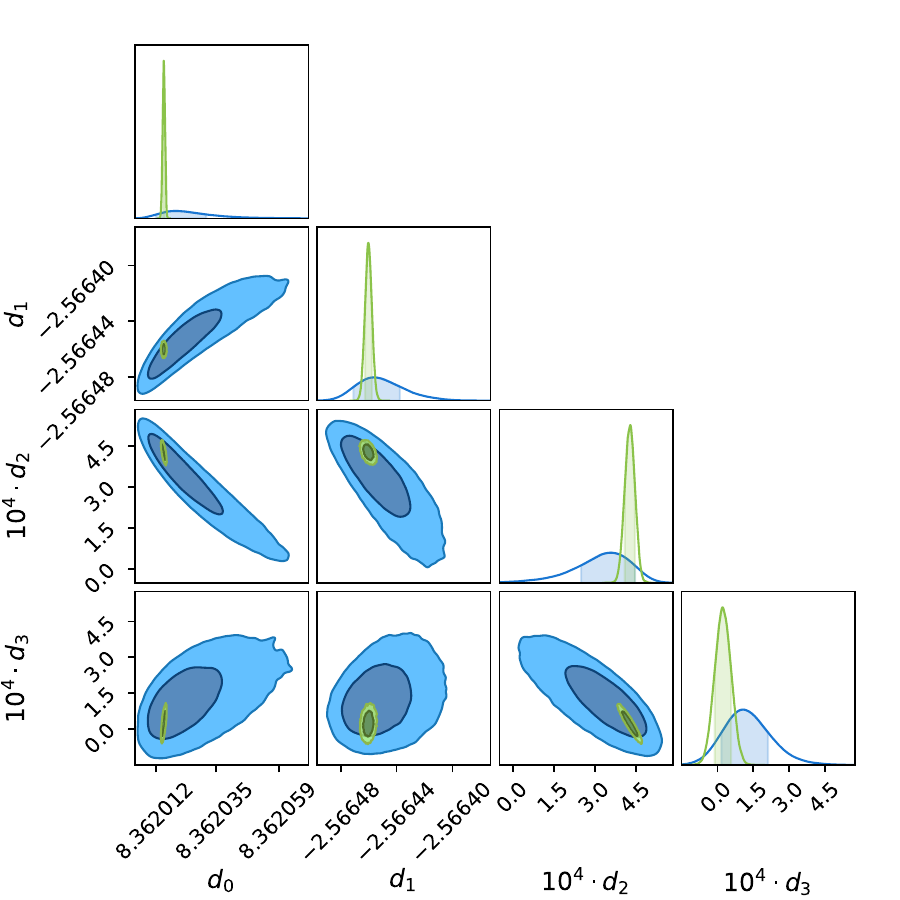}
    \caption{Posterior densities of the $N_\mathrm{poly}=4$ log-polynomial inference in a single sky bin for 100 hours of GMD-only observations (green) and GMD plus 21-cm monopole observations (blue). In the latter case, we have marginalized over the 21-cm signal parameters.}
    \label{fig:monopole_covariance_corner}
\end{figure}

\begin{figure}
    \centering
    \includegraphics[width=\columnwidth]{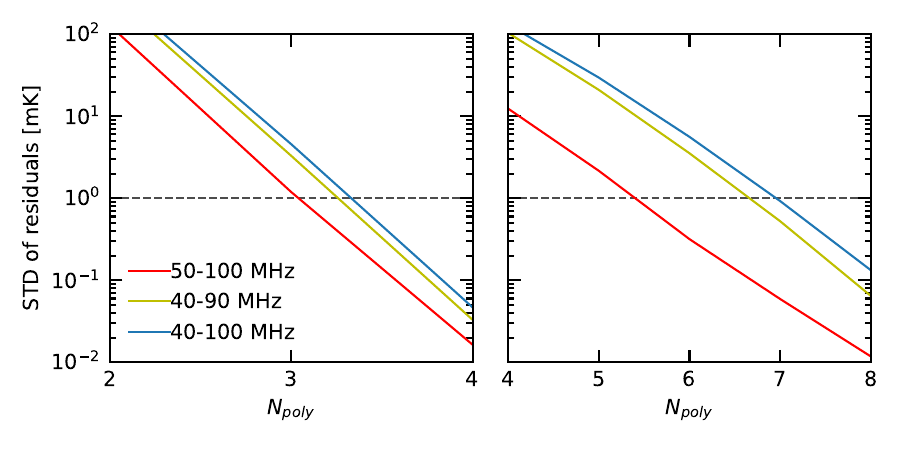}
    \caption{The standard deviation of the residuals of fitting a polynomial foreground model to the GMD (left) and GSMA (right), for varying polynomial order. The fit to three different frequency ranges is considered. Black dashed lines indicate the approximate magnitude of the dipole signal, providing a lower bound for the polynomial order required to fit the foregrounds sufficiently well to observe the dipole signal in the residuals.}
    \label{fig:binwise_stderr_onebin}
\end{figure}

To increase complexity, we consider 21-cm parameter inference in the presence of the GMD foregrounds. The typical approach of foreground-signal separation in 21-cm global experiments assumes that the spectrally smooth foregrounds may be fit using spectrally smooth log-polynomial functions, while the non-spectrally smooth 21-cm signal is fit using a Gaussian-like model. Usually, this involves time-averaging all timeseries data and solely working in the frequency domain. Since inferring the dipole direction involves at least three time bins and two antennae, we extend this method by fitting a separate log-polynomial to the observed temperature in each bin of each antenna, given by
\begin{equation}
    \log P^{N_\mathrm{poly}}_{j}(\nu) = \sum_{q=0}^{N_\mathrm{poly}-1} d_{q,j} \left[ \log(\nu/60 \, \mathrm{MHz}) \right]^q \,,
    \label{eq:binwise}
\end{equation}
where $j$ is the bin number, $N_\mathrm{poly}$ is the dimension of the log-polynomial and $d_{q,j}$ are a set of $N_\mathrm{poly}\times N_\mathrm{bins}$ independent coefficients. The noisy simulated observations in each bin, $T^\mathrm{obs}_j$, are fit simultaneously across all bins by the function $P^{N_\mathrm{poly}}_{j} +T^\mathrm{obs}_{21,j}$. Where we refer to fitting the log-polynomial to observations of a single sky bin, we drop the bin index $j$.

This method is simple and can work well for spectrally smooth foregrounds in the absence of beam achromaticity or unknown systematic errors. \cite*{2015:BernardiMcQuinnGreenhill} show that a $7^\mathrm{th}$ order polynomial can extract the 21-cm monopole if the experiment has a chromatic beam, while a $5^\mathrm{th}$ order polynomial is sufficient in the achromatic case. While their exact foreground model is different, \cite{2018:BowmanRogersMonsalve} use a 5-term, smooth foreground model with the same methodology for one of the two EDGES data analysis pipelines. 

An issue inherent to this method and well known in the context of 21-cm global signal extraction is the covariance between the foreground polynomial and the 21-cm monopole parameters. To demonstrate this effect, we simulate the observations of both the GMD sky alone, and the GMD sky with an added 21-cm monopole component, both for a single sky bin with an integration time of 100 hours. The observations are fitted by $P^4$, and $P^4 + T_\mathrm{mon}$ respectively, and the marginalised foreground parameter posteriors are shown in \cref{fig:monopole_covariance_corner}. The uncertainty in foreground parameters dramatically increases with the addition of the 21-cm monopole, driven by the degeneracy between the two model sectors.

This methodology suffers from additional issues, namely, \cite{2021:AnsteydeLeraAcedoHandley} found that the achromaticity in the antenna beam can couple to either the shape of the galaxy or its spectral features, leading to false troughs larger than the predicted 21-cm signal.

Improvements on this method have been proposed, such as \textit{maximally smooth function} analysis by \cite{2019:SinghSubrahmanyan}. This method ensures that there are no turning points in the derivatives of \cref{eq:binwise}. Since the 21-cm signal features turning points in its derivatives, maximally smooth functions cannot fit away the signal's main features. This advantage comes at the cost of less flexibility; any instrumental artefacts that have not been accounted for and are less smooth than the foregrounds cannot be fitted away by the polynomial, so will be confused with the signal. Various other methods which capture this flexibility while remaining independent to the 21-cm monopole degrees of freedom have been suggested, such as the SVD analysis of (\citealt{2018:TauscherRapettiBurns}; \citealt*{2020a:TauscherRapettiBurns}), which uses the spatial uniformity of the monopole signal to discard the requirement for an assumed signal shape.

However, the consideration of extra complications such as achromaticity, systematic errors and ionospheric effects are beyond the scope of this paper, in which we assume good knowledge of the nature of the source of the data and the components present within it. We aim to grasp the basic feasibility of the 21-cm dipole's observation, so we take the simplest reasonable foreground model.
\newline

We find an estimate for the value of $N_\mathrm{poly}$ necessary for the foreground polynomial model to encapsulate either the GMD or the GSMA foreground behaviour by calculating the standard deviation of the residuals to the fit for different values of $N_\mathrm{poly}$. We measure the sky temperature for a single all-day time bin, using a total integration time of $10^7$ hours to avoid the result being noise-limited. This integration time results in a mean noise level of 0.01 mK, which is completely sub-dominant to the scale of the 21-cm dipole. As shown in \cref{fig:binwise_stderr_onebin}, the residuals decrease with polynomial order roughly as a power law, with the GSMA fit requiring greater $N_\mathrm{poly}$ than the GMD. It must be emphasised that these results are strictly a lower bound of the polynomial order required to recover a signal without bias when coupled with a signal model.

Notably, the frequency range chosen to observe has a large impact on the goodness-of-fit, with the polynomial model especially struggling to fit lower frequency components. This is to be expected; the residuals of the log-polynomial fit to the full foreground temperature always diverge for low frequencies. To see this, assume that a non-truncated log-polynomial, with a certain choice of parameters, can perfectly fit the GMD or GSMA. The residuals obtained by subtracting a truncated log-polynomial with any choice of parameters $d_q$, from this non-truncated polynomial, will always diverge as $\nu\rightarrow0$.

\subsection{Single-antenna inference}
\label{sec:SIMPLE FOREGROUND MODEL INFERENCE - Single Antenna Inference}
\begin{figure}
    \centering
    \includegraphics[width=\columnwidth]{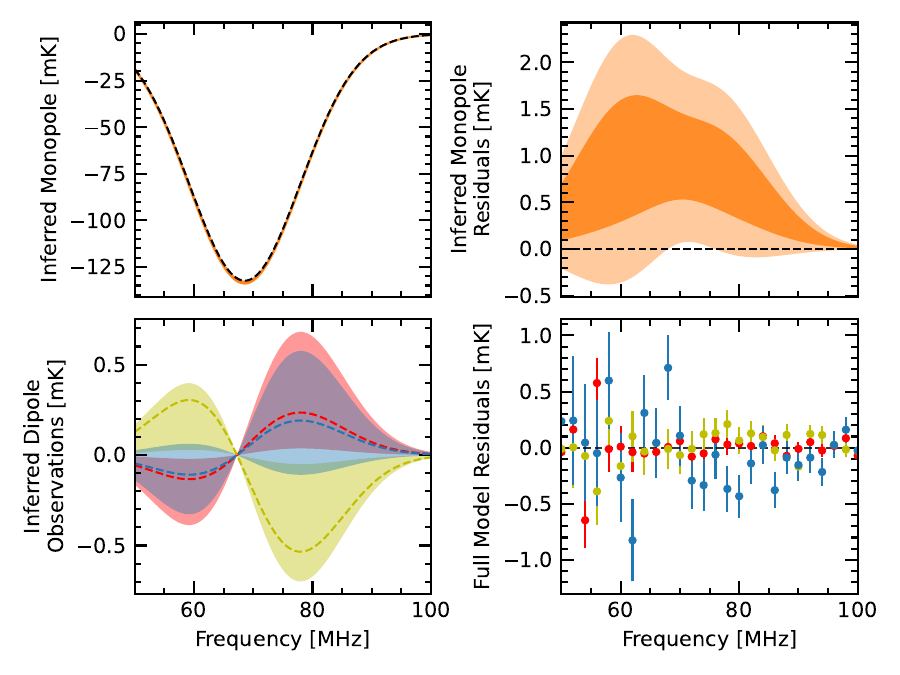}
    \caption{Results from the $N_\mathrm{poly}=4$ inference of single antenna observations of the GMD and 21-cm monopole and dipole sky across three time bins, with an integration time of $10^5$ hours. \textit{Top-left:} fiducial monopole (black, dashed) and the $1\sigma$ and $2\sigma$ regions of the inferred monopole (orange). \textit{Top-right:} residuals of the fiducial monopole curve to the inferred monopole. \textit{Bottom-left:} fiducial dipole signals (dashed) measured in each of the three time bins, along with the corresponding $1\sigma$ regions of the inferred dipole. \textit{Bottom-right:} mock data residuals of the full model compared to zero (dashed).}
    \label{fig:mondip_dip_N4_inference}
\end{figure}

\begin{figure}
    \centering
    \includegraphics[width=\columnwidth]{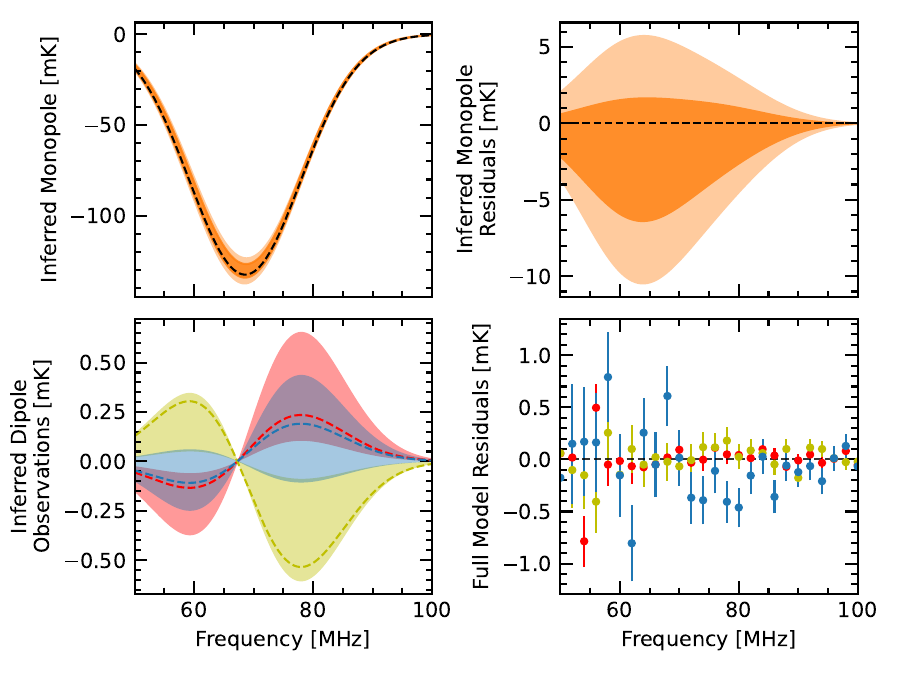}
    \caption{As for \cref{fig:mondip_dip_N4_inference}, with $N_\mathrm{poly}=5$.}
    \label{fig:mondip_dip_N5_inference}
\end{figure}

\begin{figure}
    \centering
    \includegraphics[width=\columnwidth]{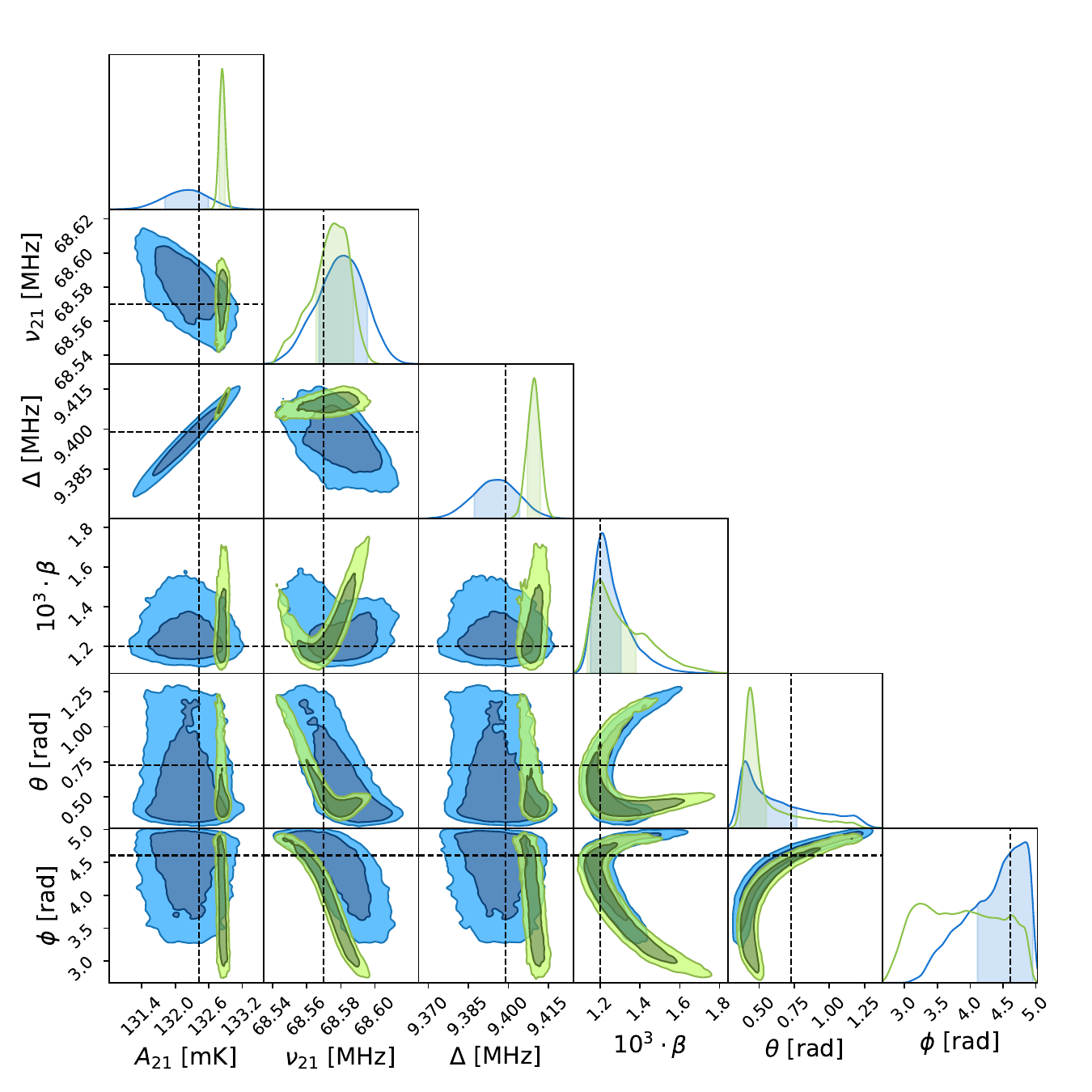}
    \caption{Marginal posteriors of the 21-cm monopole and dipole parameters, inferred by observations of the GDM and 21-cm monopole and dipole sky in 3 time bins and $10^5$ hours of integration time. A foreground log-polynomial with $N_\mathrm{poly}=4$ (green) and $N_\mathrm{poly}=5$ (blue) is used to fit the foregrounds in each bin.}
    \label{fig:fgmondip_Npoly_bias}
\end{figure}
We now consider the inference of the 21-cm parameters in the presence of GMD foregrounds, from the observations of a single dipole antenna experiment situated at the MRO site, with a total integration time of $10^5$ hours. For simplicity, the observations are binned across three time bins. We fit the model $P^4_{j}(\nu) + T^\mathrm{obs}_{21,j}(\nu)$ for $j\in\{1, 2, 3\}$
to these simulated observations. The results are shown in \cref{fig:mondip_dip_N4_inference}. The top-left panel shows the fiducial 21-cm monopole signal (black, dashed). This is compared to the orange 1$\sigma$ and 2$\sigma$ regions, generated by evaluating $T_\mathrm{mon}$ at a random sample of 1000 points drawn from the marginalised monopole parameter posterior distribution. The marginalised posterior distribution clearly shows the presence of a monopole with a size of the correct magnitude. The top-right panel shows the residuals after the 1$\sigma$ and 2$\sigma$ regions are subtracted from the fiducial monopole of the top-left panel. These illustrate the uncertainty region around the inferred monopole. While it is less visible in the prior panel, the posterior distribution is clearly biased away from the fiducial line.

The bottom-left panel shows the fiducial 21-cm dipole temperatures observed in each of the three time bins (dashed). This is compared to the 1$\sigma$ regions corresponding to each bin, generated by evaluating the observed dipole temperature (the second term of~equation~\ref{eq:observed_mondip}) at a random sample of 1000 points drawn from the marginalised 21-cm parameter posterior distribution. This is essentially the same information for the dipole that the top-left panel shows for the monopole. The yellow and red bins show weak detections of the dipole signal, as their $1\sigma$ regions are inconsistent with zero. The blue bin detects a signal consistent with zero to $1\sigma$, likely due to the larger noise values associated with the bin, which points towards the galactic centre.

The bottom-right panel shows the residuals after the full model, evaluated at the mean posterior point, has been subtracted from the mock data. The mean point has been chosen to minimise the chi-squared statistic of the residuals.
The larger noise in the blue bin is apparent. The residuals are consistent with zero.
\newline

\Cref{fig:mondip_dip_N5_inference} is the same as \cref{fig:mondip_dip_N4_inference}, corresponding to the parameter inference of the same observed data, with a $N_\mathrm{poly}=5$ foreground polynomial instead of an $N_\mathrm{poly}=4$ polynomial. Again, the model infers a 21-cm monopole signal with the correct magnitude and shape, which this time is not biased away from the fiducial monopole. However, the 21-cm monopole uncertainty region is much larger than in the previous case, despite the same data noise. Both the full model residuals and the 21-cm dipole uncertainty regions are almost identical to the $N_\mathrm{poly}=4$ case, with a weak $1\sigma$ detection of the 21-cm dipole present in the yellow and the red bins.
\newline

Increasing the foreground polynomial order both decreases the bias and increases the variance of the inferred 21-cm monopole parameters. These effects are seen in the marginalised 21-cm monopole parameter posteriors shown in \cref{fig:fgmondip_Npoly_bias}. The biased inference of the monopole posteriors for $N_\mathrm{poly}=4$ and the broad monopole posteriors of $N_\mathrm{poly}=5$ are caused by the interaction between the foreground polynomial and 21-cm Gaussian model sectors alone, which is made clear by the similarity between the inferred 21-cm dipole parameter posteriors of the two cases. The $N_\mathrm{poly}=4$ polynomial partially uses the Gaussian monopole degree of freedom to fit the foregrounds, biasing the 21-cm monopole parameters. Meanwhile, the $N_\mathrm{poly}=5$ polynomial has the complexity required to capture the foregrounds. However, the freedom this adds to the model can be mistaken for a Gaussian, broadening the variance of the inferred 21-cm monopole parameters.

Nevertheless, this degeneracy is limited to affecting the foreground and monopole model sectors, while the inferred 21-cm dipole and total model residuals remain reasonably unaffected by the change in foreground model. The results of the following sections should be treated with caution with this degeneracy in mind. However, a less degenerate foreground modelling approach will only improve estimates for the feasibility of signal detection. Therefore, we believe it is safe to assume that rough upper bounds may be set using this method.

\subsection{Multi-antenna inference}
\label{sec:multi-antenna inference}
\begin{figure}
    \centering
    \includegraphics[width=\columnwidth]{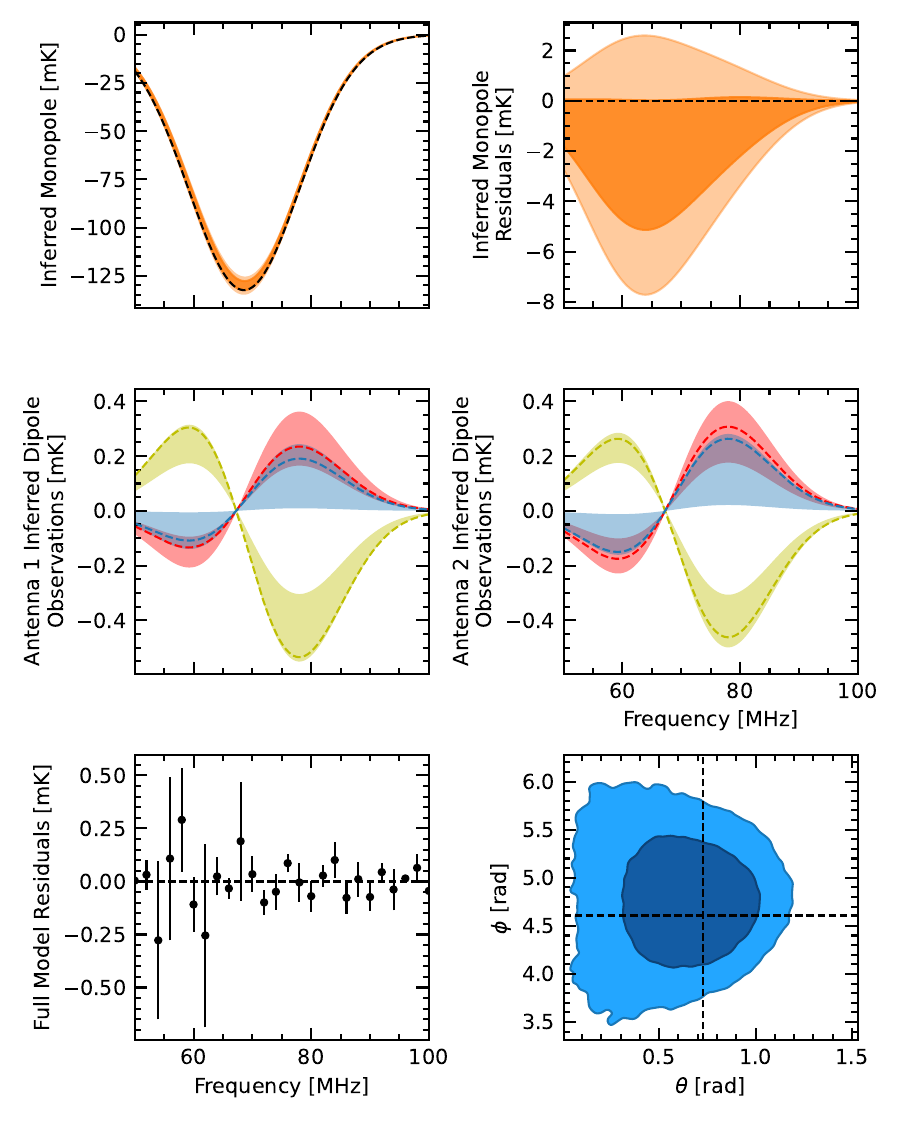}
    \caption{Results from the $N_\mathrm{poly}=5$ inference of two-antenna observations of the GMD and 21-cm monopole and dipole sky across three time bins and an integration time $10^5$ hours per antenna. Panels are as for \cref{fig:mondip_dip_N4_inference}, with the following additional panels -- \textit{Middle-right:} antenna 2's inferred dipole observations. \textit{Bottom-right:} the marginalised posterior distribution of the 21-cm dipole direction parameters, with the fiducial value shown (dashed lines). For clarity, the full model residuals are averaged across time bins in this figure, producing a single value for each frequency bin.}
    \label{fig:twoant_mondip_1e5_inference}
\end{figure}

\begin{figure}
    \centering
    \includegraphics[width=\columnwidth]{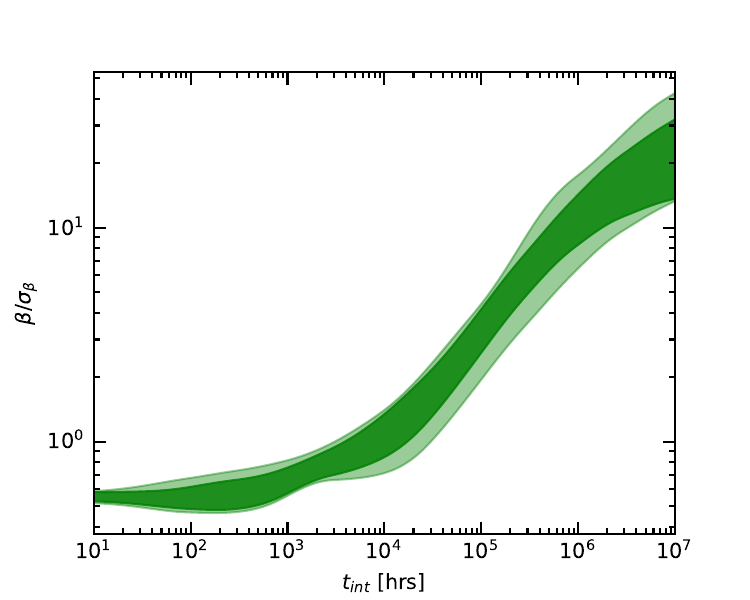}
    \caption{The detection SNR of the 21-cm dipole magnitude parameter $\beta$ as for~\cref{fig:2ant_beta_SNR}, but in the presence of GMD foregrounds. The foregrounds are fit with an $N_\mathrm{poly}=5$ log-polynomial. Only two antennae with three time bins each are considered here.}
    \label{fig:snr_beta_fg}
\end{figure}

As for the single-antenna inference of~\cref{sec:fg_free_inf}, the 21-cm dipole parameter posteriors inferred in~\cref{sec:SIMPLE FOREGROUND MODEL INFERENCE - Single Antenna Inference} are degenerate, and the dipole location is not localised to a single point in the sky. To investigate the possibility of fully leveraging the CMB dipole to validate the 21-cm signal detection in the presence of simple foregrounds, we must again consider the observations from two latitude-separated dipole antennae.

As in \cref{sec:two_antenna}, we extend the observation of the GMD and 21-cm sky to two antennae, with antenna 1 at latitude $-26.7\degr$ and antenna 2 at latitude $26.7\degr$. Each antenna's observations are again binned in three time bins, and each bin's observations are fit by the model $P^5_{j}(\nu) + T^\mathrm{obs}_{21,j}(\nu)$ for $j\in\{1, \ldots, 6\}$. The total model dimension is $N_\mathrm{dof}=36$. The inference is showcased in \cref{fig:twoant_mondip_1e5_inference}, which is laid out as \cref{fig:mondip_dip_N4_inference,fig:mondip_dip_N5_inference} are, with the addition of a 21-cm dipole panel for antenna 2, and a panel showing the marginalised posterior distribution of the 21-cm dipole direction. The dipole panels are colour-coded in the same way as~\cref{fig:breaking_degeneracy_projplot}, with time bins of both antennae corresponding to the same observation longitudes having the same colour. The two 21-cm dipole panels appear similar, as the fiducial dipole location is approximately equidistant to the two observation tracks, and the two observation tracks' time bins are aligned in longitude. For clarity, the full model residuals here are represented as the mean and standard deviation of the six time bin temperatures within each frequency bin. Given that the points in each bin are correlated, this only serves to give an idea of the spread of temperatures in each bin.

As for the single-antenna $N_\mathrm{poly}=5$ inference, the inferred monopole is not biased, but the uncertainty associated with it is again significantly larger than the data noise. This is to be expected; antenna 1 models the foregrounds it observes independently to antenna 2, so antenna 2's data cannot be used to `help' distinguish the foregrounds and 21-cm monopole temperatures that antenna 1 observes.

The dipole inference is significantly improved over the single-antenna case, with more significant dipole signal detections in the yellow and red bins of each antenna. Even the blue bins of both antennae weakly detect the dipole signal to $1\sigma$. This improvement is not seen when we run the $N_\mathrm{poly}=5$ inference on single-antenna mock observations with double the integration time used in
\cref{sec:SIMPLE FOREGROUND MODEL INFERENCE - Single Antenna Inference} ($2\times10^5$ hours). This indicates that the improvement is not caused by the effective doubling of integration time that the second antenna provides. Rather, it results from the broken 21-cm dipole parameter degeneracy.

To estimate the integration time required to detect the 21-cm dipole signal in the presence of foregrounds, we carry out the same SNR analysis for the inferred value of $\beta$ as in~\cref{sec:two_antenna}. We consider the case with three time bins and two antennae only, since any more was seen to only marginally change the result in the foreground-free analysis. The SNR rises more slowly than for the foreground-free case, reaching a confident detection threshold for $\sim 10^5 - 10^6$ hours. For much higher integration times, the steady SNR increase begins to trail off, possibly indicating a threshold where the inference becomes totally model-limited.

\section{Dipole Detection in Practice}
\label{sec:realistic_foreground_model_inference}
\begin{figure}
    \centering
    \includegraphics[width=\columnwidth]{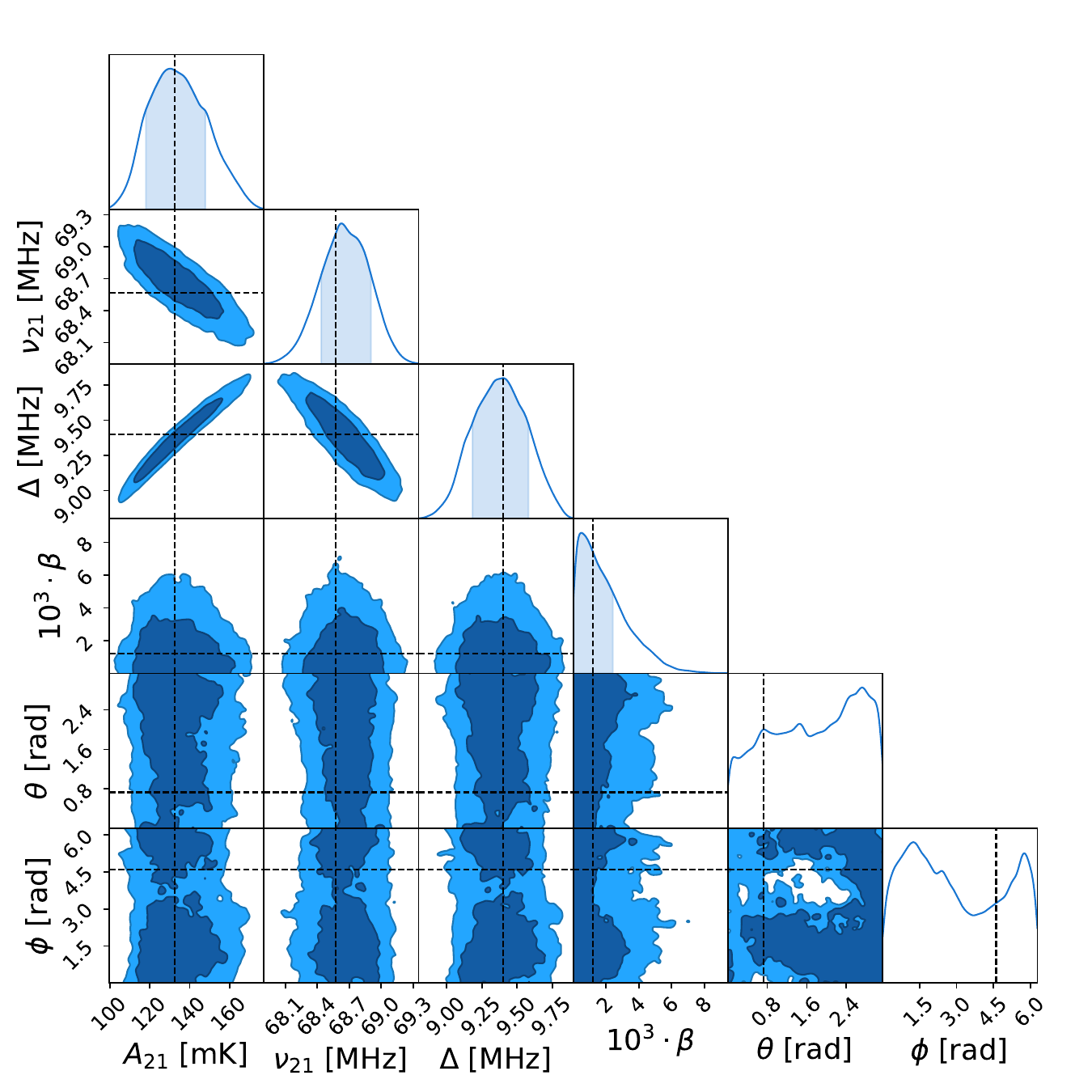}
    \caption{Marginal 21-cm posteriors inferred from single-antenna observations of the GSMA and 21-cm sky. The observations have been made in three time bins with an integration time of $10^5$ hours. An $N_\mathrm{poly}=8$ log-polynomial is used to fit the foregrounds in each bin.}
    \label{fig:gsma_dip_N8_lowint_corner}
\end{figure}

\begin{figure}
    \centering
    \includegraphics[width=\columnwidth]{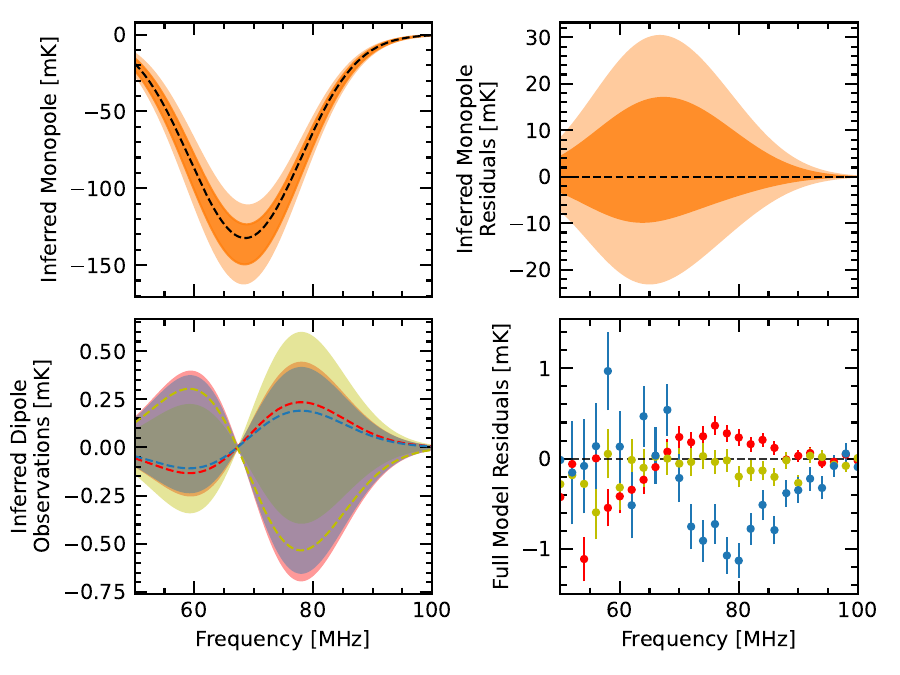}
    \caption{As for \cref{fig:mondip_dip_N4_inference}, with $N_\mathrm{poly}=8$ and the GMD foregrounds replaced by the GSMA foregrounds.}
    \label{fig:gsma_dip_N8_lowint_inference}
\end{figure}

Thus far, we have shown that a dipole detection that corroborates a 21-cm all-sky detection could be possible with the data from at least two latitude-separated drift-scan dipole antenna experiments. Unfortunately, the integration times we estimate for this, even in the absence of foregrounds, are prohibitively large. However, in \cref{sec:two_antenna}, we have shown that additional data from more than two antennae scales the detection SNR of the dipole magnitude with the extra integration time that the additional antennae provide. Assuming a perfect foreground removal method, a global network of 20 such antennae could reasonably detect the 21-cm monopole and dipole signals concurrently, with a much more reasonable integration time of $10^3$ hours. Meanwhile, our results in the presence of the GMD foregrounds indicate that significantly higher total integration times may be required. Assuming the same relationship holds in this case between the detection SNR and the number of antennae, we estimate that a network of $\sim$200 global antennae could be utilised to achieve a $4\sigma$ dipole detection with $10^3$ hours of integration time.

A caveat to this is the somewhat simple GMD foreground component and the polynomial modelling used thus far in this work. When mock observations of the full GSMA foregrounds and 21-cm sky are fit using a foreground polynomial and 21-cm model, the result is poor. With one antenna, three time bins and $N_\mathrm{poly}=8$, the model succeeds in inferring the 21-cm monopole parameters without bias, but fails to convincingly infer the dipole parameters (\cref{fig:gsma_dip_N8_lowint_corner}). The posterior distribution of $\beta$ is running up against the lower prior bound, and tails off for larger values. The model inference is effectively choosing values of $\beta$ that make the 21-cm dipole sector small compared to the residuals of the foreground and 21-cm monopole sectors. This results in an unconstrained 21-cm dipole direction.

\Cref{fig:gsma_dip_N8_lowint_inference} shows the results of this inference in the same format as \cref{fig:mondip_dip_N4_inference,,fig:mondip_dip_N5_inference}. The 21-cm monopole is detected, and is not biased away from the fiducial line. However, the inferred monopole's variance is much larger than the monopole inferred when an $N_\mathrm{poly}=5$ polynomial is used to fit the GMD foregrounds (\cref{sec:SIMPLE FOREGROUND MODEL INFERENCE - Single Antenna Inference}). This is likely caused by the extra foreground degrees of freedom, which are covariant with the Gaussian signal model. As expected from the unconstrained 21-cm dipole parameters, the dipole observations carry no information. Additionally, the full model residuals are biased.

Given the much greater spectral complexity of the GSMA over the GDM as discussed in \cref{sec:Foreground Sky Models}, a foreground polynomial with a greater number of terms is required to fit these observations. This raises two issues with the foreground modelling strategy used in this work. Namely, the foreground model dimension is $N_\mathrm{poly}\times N_\mathrm{bins}\times N_\mathrm{antennae}$, so any increase in foreground polynomial order corresponds to a large increase in model dimension. {\small EMCEE} is not made to handle inference with large numbers of parameters. Leveraging other packages such as {\small PYMC} \citep{2023:pymc}, for Hamiltonian Monte-Carlo sampling, or {\small POLYCHORD} \citep*{2015a:HandleyHobsonLasenby,2015b:HandleyHobsonLasenby}, for nested sampling, may provide a solution, which we leave to future work.

Secondly, since the foreground model fits observations in each time bin independently, it has no access to the information carried by the spatial variation of the foregrounds. This strategy risks overfitting the data. To fit GSMA observations, the amount of freedom that the foreground model must be given means that it will likely be able to completely fit the spectral form of the 21-cm dipole in each of the observation bins.

To alleviate this, a more sophisticated foreground model which takes the spatial structure of the foreground sky into account may be utilised. Multiple such modelling approaches have been developed for global 21-cm experiments. For instance, \cite{2021:AnsteydeLeraAcedoHandley, 2023:AnsteydeLeraAcedoHandley} split the sky into $N$ separate regions, each with their own amplitude and power law index. \cite{2018:TauscherRapettiBurns}, \cite{2020:RapettiTauscherMirocha} and \cite*{2020b:TauscherRapettiBurns} develop a formalism that leverages the differing spatial variation of the foreground and 21-cm signals across different antenna pointings. \cite{2020a:TauscherRapettiBurns} include a discussion on how to apply this formalism to also extract the 21-cm dipole. An alternate strategy that may be explored is modelling the foregrounds using a set of frequency varying spherical harmonics with coefficients $a_{lm}(\nu)$. The spectral form of the $a_{lm}$ would have to be chosen carefully to mitigate the risk of fitting out the 21-cm signal. All of these suggested foreground models' dimensions do not scale with the number of antennae used. It is possible that by using any of these strategies, data from additional antennae would increase the 21-cm dipole detection SNR by more than just the extra integration time they add, as each antenna would work together to constrain the same foreground sky.

Clearly, real-world data requires more sophisticated foreground modelling than the log-polynomial fitting used in this work. While more realistic foregrounds act to increase the integration times required for a dipole detection, these may be counteracted by better-suited foreground modelling and observation strategies.

\section{Conclusion}
\label{sec:conclusion}
In this work, we have investigated the minimum requirements for drift-scan dipole antenna experiments to confidently detect the 21-cm dipole, with the aim to verify a global 21-cm signal detection. We have found that the data necessary to localise the dipole position and magnitude consists of the observed temperatures by at least two antennae situated at different latitudes, with at least three time bins per antenna. This finding suggests that global 21-cm experiments based in the Northern hemisphere, such as MIST, may significantly contribute to the search of the 21-cm dipole. An added benefit is that an antenna situated in the Northern hemisphere faces away from the radio-bright galactic centre. 

In the absence of foregrounds, instrumental or ionospheric effects, we find that the total integration time required to detect the dipole to a few $\sigma$ is $\sim 10^4$ hours. This estimate indicates that even if EDGES and SARAS-3 data were combined, with a perfect foreground removal strategy, we do not expect a dipole detection to be possible. However, a global network of 20 such antennae could achieve this detection in $\sim 10^3$ hours. For the more realistic case of detecting the 21-cm dipole in mock observations contaminated by simple foregrounds, we generalised the standard log-polynomial model used for 21-cm global signal foreground fitting to the case of multiple time bins and antennae. The total integration time required for a two-antenna experiment to detect the dipole to a few $\sigma$ rose to $\sim 10^5 - 10^6$ hours, which we estimate could be achieved within $\sim 10^3$ hours with a global network of 200 antennae.

When the 21-cm signal is contaminated by realistic foregrounds, the log-polynomial strategy fails to recover the 21-cm dipole for any reasonable integration time. The covariance between the foreground and signal sectors of the log-polynomial model is known to be a major source of uncertainty in global 21-cm signal extraction. A multitude of improvements to the strategy exist in the literature, some of which may be applied to the 21-cm dipole context. We motivate the use of two such models, both of which remove the need to bin timeseries data, and could make global antenna networks more powerful by having the data from each antenna work together to constrain the full foreground model. The application of such foreground models is beyond the scope of this work, in which we seek basic insights into the possibility of detecting the 21-cm dipole with global drift-scan antenna experiments. 

Detection of the 21-cm dipole is clearly a challenging proposition. Central to this effort is a need to better exploit not only spectral, but also spatial, information from 21-cm global experiments. Further consideration of how best to exploit time stream data from widely separated global experiments seems warranted. More detailed work would feature more sophisticated foreground modelling, and would take beam chromaticity and ionospheric effects into account. Attempts to measure the 21-cm dipole and other low-order multipoles may prove highly complementary to existing monopole detection efforts.

\section*{Acknowledgements}
YI acknowledges the support of a Science and Technology Facilities Council studentship [grant number ST/W507519/1]. Some of the results in this paper have been derived using the {\small HEALPY} and {\small HEALPIX} package.

\section*{Data Availability}
The data underlying this article will be shared on reasonable request to the corresponding author.



\bibliographystyle{mnras}
\bibliography{bibliography}

\begin{thebibliography}{}
\makeatletter
\relax
\def\mn@urlcharsother{\let\do\@makeother \do\$\do\&\do\#\do\^\do\_\do\%\do\~}
\def\mn@doi{\begingroup\mn@urlcharsother \@ifnextchar [ {\mn@doi@}
  {\mn@doi@[]}}
\def\mn@doi@[#1]#2{\def\@tempa{#1}\ifx\@tempa\@empty \href
  {http://dx.doi.org/#2} {doi:#2}\else \href {http://dx.doi.org/#2} {#1}\fi
  \endgroup}
\def\mn@eprint#1#2{\mn@eprint@#1:#2::\@nil}
\def\mn@eprint@arXiv#1{\href {http://arxiv.org/abs/#1} {{\tt arXiv:#1}}}
\def\mn@eprint@dblp#1{\href {http://dblp.uni-trier.de/rec/bibtex/#1.xml}
  {dblp:#1}}
\def\mn@eprint@#1:#2:#3:#4\@nil{\def\@tempa {#1}\def\@tempb {#2}\def\@tempc
  {#3}\ifx \@tempc \@empty \let \@tempc \@tempb \let \@tempb \@tempa \fi \ifx
  \@tempb \@empty \def\@tempb {arXiv}\fi \@ifundefined
  {mn@eprint@\@tempb}{\@tempb:\@tempc}{\expandafter \expandafter \csname
  mn@eprint@\@tempb\endcsname \expandafter{\@tempc}}}

\bibitem[\protect\citeauthoryear{{Anstey}, {de Lera Acedo}  \&
  {Handley}}{{Anstey} et~al.}{2021}]{2021:AnsteydeLeraAcedoHandley}
{Anstey} D.,  {de Lera Acedo} E.,   {Handley} W.,  2021, \mn@doi [\mnras]
  {10.1093/mnras/stab1765}, \href
  {https://ui.adsabs.harvard.edu/abs/2021MNRAS.506.2041A} {506, 2041}

\bibitem[\protect\citeauthoryear{{Anstey}, {de Lera Acedo}  \&
  {Handley}}{{Anstey} et~al.}{2023}]{2023:AnsteydeLeraAcedoHandley}
{Anstey} D.,  {de Lera Acedo} E.,   {Handley} W.,  2023, \mn@doi [\mnras]
  {10.1093/mnras/stad156}, \href
  {https://ui.adsabs.harvard.edu/abs/2023MNRAS.520..850A} {520, 850}

\bibitem[\protect\citeauthoryear{{Barkana}}{{Barkana}}{2018}]{2018:Barkana}
{Barkana} R.,  2018, \mn@doi [\nat] {10.1038/nature25791}, 555, 71

\bibitem[\protect\citeauthoryear{Bernardi, McQuinn  \& Greenhill}{Bernardi
  et~al.}{2015}]{2015:BernardiMcQuinnGreenhill}
Bernardi G.,  McQuinn M.,   Greenhill L.~J.,  2015, \mn@doi [\apj]
  {10.1088/0004-637x/799/1/90}, 799, 90

\bibitem[\protect\citeauthoryear{{Bowman}, {Rogers}, {Monsalve}, {Mozdzen}  \&
  {Mahesh}}{{Bowman} et~al.}{2018}]{2018:BowmanRogersMonsalve}
{Bowman} J.~D.,  {Rogers} A. E.~E.,  {Monsalve} R.~A.,  {Mozdzen} T.~J.,
  {Mahesh} N.,  2018, \mn@doi [\nat] {10.1038/nature25792}, \href
  {https://ui.adsabs.harvard.edu/abs/2018Natur.555...67B} {555, 67}

\bibitem[\protect\citeauthoryear{{Brandenberger}, {Cyr}  \&
  {Shi}}{{Brandenberger} et~al.}{2019}]{2019:BrandenbergerCyrShi}
{Brandenberger} R.,  {Cyr} B.,   {Shi} R.,  2019, \mn@doi [\jcap]
  {10.1088/1475-7516/2019/09/009}, \href
  {https://ui.adsabs.harvard.edu/abs/2019JCAP...09..009B} {2019, 009}

\bibitem[\protect\citeauthoryear{{Cohen}, {Fialkov}, {Barkana}  \&
  {Lotem}}{{Cohen} et~al.}{2017}]{2017:CohenFialkovBarkana}
{Cohen} A.,  {Fialkov} A.,  {Barkana} R.,   {Lotem} M.,  2017, \mn@doi [\mnras]
  {10.1093/mnras/stx2065}, \href
  {https://ui.adsabs.harvard.edu/abs/2017MNRAS.472.1915C} {472, 1915}

\bibitem[\protect\citeauthoryear{{Deshpande}}{{Deshpande}}{2018}]{2018:Deshpande}
{Deshpande} A.~A.,  2018, \mn@doi [\apjl] {10.3847/2041-8213/aae318}, \href
  {https://ui.adsabs.harvard.edu/abs/2018ApJ...866L...7D} {866, L7}

\bibitem[\protect\citeauthoryear{{Dowell} \& {Taylor}}{{Dowell} \&
  {Taylor}}{2018}]{2018:DowellTaylor}
{Dowell} J.,  {Taylor} G.~B.,  2018, \mn@doi [\apjl]
  {10.3847/2041-8213/aabf86}, \href
  {https://ui.adsabs.harvard.edu/abs/2018ApJ...858L...9D} {858, L9}

\bibitem[\protect\citeauthoryear{{Fixsen} et~al.,}{{Fixsen}
  et~al.}{2011}]{2011:FixsenKogutLevin}
{Fixsen} D.~J.,  et~al., 2011, \mn@doi [\apj] {10.1088/0004-637X/734/1/5},
  \href {https://ui.adsabs.harvard.edu/abs/2011ApJ...734....5F} {734, 5}

\bibitem[\protect\citeauthoryear{{Foreman-Mackey}, {Hogg}, {Lang}  \&
  {Goodman}}{{Foreman-Mackey} et~al.}{2013}]{2013:Foreman-MackeyHoggLang}
{Foreman-Mackey} D.,  {Hogg} D.~W.,  {Lang} D.,   {Goodman} J.,  2013, \mn@doi
  [\pasp] {10.1086/670067}, \href
  {https://ui.adsabs.harvard.edu/abs/2013PASP..125..306F} {125, 306}

\bibitem[\protect\citeauthoryear{{Fraser} et~al.,}{{Fraser}
  et~al.}{2018}]{2018:FraserHektorHutsi}
{Fraser} S.,  et~al., 2018, \mn@doi [Phys. Lett. B]
  {10.1016/j.physletb.2018.08.035}, \href
  {https://ui.adsabs.harvard.edu/abs/2018PhLB..785..159F} {785, 159}

\bibitem[\protect\citeauthoryear{{Furlanetto}, {Oh}  \& {Briggs}}{{Furlanetto}
  et~al.}{2006}]{2006:FurlanettoOhBriggs}
{Furlanetto} S.~R.,  {Oh} S.~P.,   {Briggs} F.~H.,  2006, \mn@doi [\physrep]
  {10.1016/j.physrep.2006.08.002}, \href
  {https://ui.adsabs.harvard.edu/abs/2006PhR...433..181F} {433, 181}

\bibitem[\protect\citeauthoryear{{Handley}, {Hobson}  \& {Lasenby}}{{Handley}
  et~al.}{2015a}]{2015a:HandleyHobsonLasenby}
{Handley} W.~J.,  {Hobson} M.~P.,   {Lasenby} A.~N.,  2015a, \mn@doi [\mnras]
  {10.1093/mnrasl/slv047}, \href
  {https://ui.adsabs.harvard.edu/abs/2015MNRAS.450L..61H} {450, L61}

\bibitem[\protect\citeauthoryear{{Handley}, {Hobson}  \& {Lasenby}}{{Handley}
  et~al.}{2015b}]{2015b:HandleyHobsonLasenby}
{Handley} W.~J.,  {Hobson} M.~P.,   {Lasenby} A.~N.,  2015b, \mn@doi [\mnras]
  {10.1093/mnras/stv1911}, \href
  {https://ui.adsabs.harvard.edu/abs/2015MNRAS.453.4384H} {453, 4384}

\bibitem[\protect\citeauthoryear{{Hills}, {Kulkarni}, {Meerburg}  \&
  {Puchwein}}{{Hills} et~al.}{2018}]{2018:HillsKulkarniMeerburg}
{Hills} R.,  {Kulkarni} G.,  {Meerburg} P.~D.,   {Puchwein} E.,  2018, \mn@doi
  [\nat] {10.1038/s41586-018-0796-5}, \href
  {https://ui.adsabs.harvard.edu/abs/2018Natur.564E..32H} {564, E32}

\bibitem[\protect\citeauthoryear{{Hotinli} \& {Ahn}}{{Hotinli} \&
  {Ahn}}{2023}]{2023:HotinliAhn}
{Hotinli} S.~C.,  {Ahn} K.,  2023, \mn@doi [arXiv e-prints]
  {10.48550/arXiv.2305.01672}, \href
  {https://ui.adsabs.harvard.edu/abs/2023arXiv230501672H} {p. arXiv:2305.01672}

\bibitem[\protect\citeauthoryear{{Mirocha}}{{Mirocha}}{2014}]{2014:Mirocha}
{Mirocha} J.,  2014, \mn@doi [\mnras] {10.1093/mnras/stu1193}, \href
  {https://ui.adsabs.harvard.edu/abs/2014MNRAS.443.1211M} {443, 1211}

\bibitem[\protect\citeauthoryear{{Mirocha} \& {Furlanetto}}{{Mirocha} \&
  {Furlanetto}}{2019}]{2019:MirochaFurlanetto}
{Mirocha} J.,  {Furlanetto} S.~R.,  2019, \mn@doi [\mnras]
  {10.1093/mnras/sty3260}, \href
  {https://ui.adsabs.harvard.edu/abs/2019MNRAS.483.1980M} {483, 1980}

\bibitem[\protect\citeauthoryear{{Mittal} \& {Kulkarni}}{{Mittal} \&
  {Kulkarni}}{2022}]{2022:MittalKulkarni}
{Mittal} S.,  {Kulkarni} G.,  2022, \mn@doi [\mnras] {10.1093/mnras/stac1961},
  \href {https://ui.adsabs.harvard.edu/abs/2022MNRAS.515.2901M} {515, 2901}

\bibitem[\protect\citeauthoryear{{Monsalve} et~al.,}{{Monsalve}
  et~al.}{2023}]{2023:MonsalveByeSievers}
{Monsalve} R.~A.,  et~al., 2023, \mn@doi [arXiv e-prints]
  {10.48550/arXiv.2310.07741}, \href
  {https://ui.adsabs.harvard.edu/abs/2023arXiv231007741M} {p. arXiv:2310.07741}

\bibitem[\protect\citeauthoryear{{Mu{\~n}oz} \& {Loeb}}{{Mu{\~n}oz} \&
  {Loeb}}{2018}]{2018:MunozLoeb}
{Mu{\~n}oz} J.~B.,  {Loeb} A.,  2018, arXiv e-prints, \href
  {https://ui.adsabs.harvard.edu/abs/2018arXiv180210094M} {p. arXiv:1802.10094}

\bibitem[\protect\citeauthoryear{{Nambissan T.} et~al.,}{{Nambissan T.}
  et~al.}{2021}]{2021:NambissanT.SubrahmanyanSomashekar}
{Nambissan T.} J.,  et~al., 2021, arXiv e-prints, \href
  {https://ui.adsabs.harvard.edu/abs/2021arXiv210401756N} {p. arXiv:2104.01756}

\bibitem[\protect\citeauthoryear{Oriol et~al.,}{Oriol et~al.}{2023}]{2023:pymc}
Oriol A.-P.,  et~al., 2023, \mn@doi [{PeerJ} Computer Science]
  {10.7717/peerj-cs.1516}, 9, e1516

\bibitem[\protect\citeauthoryear{{Philip} et~al.,}{{Philip}
  et~al.}{2019}]{2019:PhilipAbdurashidovaChiang}
{Philip} L.,  et~al., 2019, \mn@doi [J. Astron. Instrum.]
  {10.1142/S2251171719500041}, \href
  {https://ui.adsabs.harvard.edu/abs/2019JAI.....850004P} {8, 1950004}

\bibitem[\protect\citeauthoryear{{Planck Collaboration} et~al.,}{{Planck
  Collaboration} et~al.}{2020a}]{2020:PlanckCollaborationAghanimAkramiIII}
{Planck Collaboration} et~al., 2020a, \mn@doi [\aap]
  {10.1051/0004-6361/201832909}, \href
  {https://ui.adsabs.harvard.edu/abs/2020A&A...641A...3P} {641, A3}

\bibitem[\protect\citeauthoryear{{Planck Collaboration} et~al.,}{{Planck
  Collaboration} et~al.}{2020b}]{2020:PlanckCollaborationAghanimAkramiVI}
{Planck Collaboration} et~al., 2020b, \mn@doi [\aap]
  {10.1051/0004-6361/201833910}, \href
  {https://ui.adsabs.harvard.edu/abs/2020A&A...641A...6P} {641, A6}

\bibitem[\protect\citeauthoryear{{Pospelov}, {Pradler}, {Ruderman}  \&
  {Urbano}}{{Pospelov} et~al.}{2018}]{2018:PospelovPradlerRuderman}
{Pospelov} M.,  {Pradler} J.,  {Ruderman} J.~T.,   {Urbano} A.,  2018, \mn@doi
  [\prl] {10.1103/PhysRevLett.121.031103}, \href
  {https://ui.adsabs.harvard.edu/abs/2018PhRvL.121c1103P} {121, 031103}

\bibitem[\protect\citeauthoryear{{Price} et~al.,}{{Price}
  et~al.}{2018}]{2018:PriceGreenhillFialkov}
{Price} D.~C.,  et~al., 2018, \mn@doi [\mnras] {10.1093/mnras/sty1244}, \href
  {https://ui.adsabs.harvard.edu/abs/2018MNRAS.478.4193P} {478, 4193}

\bibitem[\protect\citeauthoryear{{Pritchard} \& {Loeb}}{{Pritchard} \&
  {Loeb}}{2010}]{2010:PritchardLoeb}
{Pritchard} J.~R.,  {Loeb} A.,  2010, \mn@doi [\prd]
  {10.1103/PhysRevD.82.023006}, \href
  {https://ui.adsabs.harvard.edu/abs/2010PhRvD..82b3006P} {82, 023006}

\bibitem[\protect\citeauthoryear{{Pritchard} \& {Loeb}}{{Pritchard} \&
  {Loeb}}{2012}]{2012:PritchardLoeb}
{Pritchard} J.~R.,  {Loeb} A.,  2012, \mn@doi [Rep. Prog. Phys.]
  {10.1088/0034-4885/75/8/086901}, \href
  {https://ui.adsabs.harvard.edu/abs/2012RPPh...75h6901P} {75, 086901}

\bibitem[\protect\citeauthoryear{{Rapetti}, {Tauscher}, {Mirocha}  \&
  {Burns}}{{Rapetti} et~al.}{2020}]{2020:RapettiTauscherMirocha}
{Rapetti} D.,  {Tauscher} K.,  {Mirocha} J.,   {Burns} J.~O.,  2020, \mn@doi
  [\apj] {10.3847/1538-4357/ab9b29}, \href
  {https://ui.adsabs.harvard.edu/abs/2020ApJ...897..174R} {897, 174}

\bibitem[\protect\citeauthoryear{{Reis}, {Fialkov}  \& {Barkana}}{{Reis}
  et~al.}{2020}]{2020:ReisFialkovBarkana}
{Reis} I.,  {Fialkov} A.,   {Barkana} R.,  2020, \mn@doi [\mnras]
  {10.1093/mnras/staa3091}, \href
  {https://ui.adsabs.harvard.edu/abs/2020MNRAS.499.5993R} {499, 5993}

\bibitem[\protect\citeauthoryear{Robertson}{Robertson}{2022}]{2022:Robertson}
Robertson B.~E.,  2022, \mn@doi [ARA&A] {10.1146/annurev-astro-120221-044656},
  60, 121

\bibitem[\protect\citeauthoryear{{Rybicki} \& {Lightman}}{{Rybicki} \&
  {Lightman}}{1986}]{1986:RybickiLightman}
{Rybicki} G.~B.,  {Lightman} A.~P.,  1986, {Radiative Processes in
  Astrophysics}.
Wiley-VCH, Weinheim, Germany

\bibitem[\protect\citeauthoryear{{Sikder}, {Barkana}, {Fialkov}  \&
  {Reis}}{{Sikder} et~al.}{2023}]{2023:SikderBarkanaFialkov}
{Sikder} S.,  {Barkana} R.,  {Fialkov} A.,   {Reis} I.,  2023, \mn@doi [arXiv
  e-prints] {10.48550/arXiv.2301.04585}, \href
  {https://ui.adsabs.harvard.edu/abs/2023arXiv230104585S} {p. arXiv:2301.04585}

\bibitem[\protect\citeauthoryear{{Singal} et~al.,}{{Singal}
  et~al.}{2011}]{2011:SingalFixsenKogut}
{Singal} J.,  et~al., 2011, \mn@doi [\apj] {10.1088/0004-637X/730/2/138}, \href
  {https://ui.adsabs.harvard.edu/abs/2011ApJ...730..138S} {730, 138}

\bibitem[\protect\citeauthoryear{Singh \& Subrahmanyan}{Singh \&
  Subrahmanyan}{2019}]{2019:SinghSubrahmanyan}
Singh S.,  Subrahmanyan R.,  2019, \mn@doi [ApJ] {10.3847/1538-4357/ab2879},
  880, 26

\bibitem[\protect\citeauthoryear{{Singh} et~al.,}{{Singh}
  et~al.}{2022}]{2022:SinghJishnuSubrahmanyan}
{Singh} S.,  et~al., 2022, \mn@doi [Nature Astronomy]
  {10.1038/s41550-022-01610-5}, \href
  {https://ui.adsabs.harvard.edu/abs/2022NatAs...6..607S} {6, 607}

\bibitem[\protect\citeauthoryear{{Slosar}}{{Slosar}}{2017}]{2017:Slosar}
{Slosar} A.,  2017, \mn@doi [\prl] {10.1103/PhysRevLett.118.151301}, \href
  {https://ui.adsabs.harvard.edu/abs/2017PhRvL.118o1301S} {118, 151301}

\bibitem[\protect\citeauthoryear{{Tauscher}, {Rapetti}, {Burns}  \&
  {Switzer}}{{Tauscher} et~al.}{2018}]{2018:TauscherRapettiBurns}
{Tauscher} K.,  {Rapetti} D.,  {Burns} J.~O.,   {Switzer} E.,  2018, \mn@doi
  [\apj] {10.3847/1538-4357/aaa41f}, \href
  {https://ui.adsabs.harvard.edu/abs/2018ApJ...853..187T} {853, 187}

\bibitem[\protect\citeauthoryear{{Tauscher}, {Rapetti}  \& {Burns}}{{Tauscher}
  et~al.}{2020a}]{2020a:TauscherRapettiBurns}
{Tauscher} K.,  {Rapetti} D.,   {Burns} J.~O.,  2020a, \mn@doi [\apj]
  {10.3847/1538-4357/ab9a3f}, \href
  {https://ui.adsabs.harvard.edu/abs/2020ApJ...897..132T} {897, 132}

\bibitem[\protect\citeauthoryear{{Tauscher}, {Rapetti}  \& {Burns}}{{Tauscher}
  et~al.}{2020b}]{2020b:TauscherRapettiBurns}
{Tauscher} K.,  {Rapetti} D.,   {Burns} J.~O.,  2020b, \mn@doi [\apj]
  {10.3847/1538-4357/ab9b2a}, \href
  {https://ui.adsabs.harvard.edu/abs/2020ApJ...897..175T} {897, 175}

\bibitem[\protect\citeauthoryear{{Zheng} et~al.,}{{Zheng}
  et~al.}{2017}]{2017:ZhengTegmarkDillon}
{Zheng} H.,  et~al., 2017, \mn@doi [\mnras] {10.1093/mnras/stw2525}, \href
  {https://ui.adsabs.harvard.edu/abs/2017MNRAS.464.3486Z} {464, 3486}

\bibitem[\protect\citeauthoryear{{de Lera Acedo} et~al.,}{{de Lera Acedo}
  et~al.}{2022}]{2022:deLeraAcedodeVilliersRazavi-Ghods}
{de Lera Acedo} E.,  et~al., 2022, \mn@doi [Nat. Astron.]
  {10.1038/s41550-022-01709-9}, \href
  {https://ui.adsabs.harvard.edu/abs/2022NatAs...6..984D} {6, 984}

\bibitem[\protect\citeauthoryear{{de Oliveira-Costa}, {Tegmark}, {Gaensler},
  {Jonas}, {Landecker}  \& {Reich}}{{de Oliveira-Costa}
  et~al.}{2008}]{2008:deOliveira-CostaTegmarkGaensler}
{de Oliveira-Costa} A.,  {Tegmark} M.,  {Gaensler} B.~M.,  {Jonas} J.,
  {Landecker} T.~L.,   {Reich} P.,  2008, \mn@doi [\mnras]
  {10.1111/j.1365-2966.2008.13376.x}, \href
  {https://ui.adsabs.harvard.edu/abs/2008MNRAS.388..247D} {388, 247}

\makeatother
\end{thebibliography}

\bsp	
\label{lastpage}
\end{document}